\documentclass[11pt]{amsart}
\usepackage{comment,listings,framed,graphicx,url,xcolor,wrapfig}
\usepackage{fullpage}
\usepackage{amsmath}
\usepackage{amssymb}
\usepackage{url}
\usepackage{enumerate}
\usepackage{cases,float,bbm}
\usepackage{amssymb}
\usepackage{amsmath}
\newcommand{\tfr}{\theta_*}
\newcommand{\tfrm}{\theta_*^{(m)}}
\newcommand{\tzero}{\theta_0}
\newcommand{\tref}[1]{\theta_{ref,#1}}
\newcommand{\onechar}{\mathbf{1}}

\newcommand{\myP}[1]{P-ideal}
\newcommand{\myPR}[1]{P-realistic}
\newcommand{\chis}{\chi_*}
\newcommand{\chires}{\chi_{res}}

\newcommand{\ws}{w_*}
\newcommand{\vl}{\vartheta_{left}}
\newcommand{\vr}{\vartheta_{right}}

\newcommand{\ceff}{c_{eff}}
\newcommand{\keff}{k_{eff}}
\newcommand{\chieff}{\chi_{eff}}
\newcommand{\cCeff}{\cC_{eff}}

\newcommand{\kemp}{k_{emp}}
\newcommand{\chiemp}{\chi_{emp}}
\newcommand{\cCemp}{\cC_{emp}}

\newcommand{\cuf}{c_{uf}}
\newcommand{\cfr}{c_{fr}}

\newcommand{\degc}{^\circ C}
\newcommand{\degK}{K}
\newcommand{\mymu}{\nu}
\newcommand{\cbar}{\bar{c}}
\newcommand{\Lbar}{\bar{L}}
\newcommand{\abs}[1]{|\,#1\,|}
\newcommand{\wt}{\widetilde}

\newcommand{\myp}{\upsilon}

\newcommand{\gG}{\mathcal{G}}

\newcommand{\rass}[1]{Assumption~\ref{#1}}
\newcommand{\ralg}[1]{Algorithm~\eqref{#1}}
\newcommand{\rdef}[1]{Definition~\ref{#1}}
\newcommand{\Omegam}{\Omega^{(m)}}
\newcommand{\omegam}{\omega^{(m)}}
\newcommand{\omegap}{\omega_p}
\newcommand{\omegamp}{\omega_p^{(m)}}
\newcommand{\omegamr}{\omega_r^{(m)}}

\newcommand{\etamr}{\eta_r^{(m)}}
\newcommand{\etam}{\eta^{(m)}}
\newcommand{\satmp}{S_p^{(m)}}
\newcommand{\omegar}{\omega_r}
\newcommand{\omegaa}{\omega_a}
\newcommand{\omegag}{\omega_g}
\newcommand{\omegal}{\omega_l}
\newcommand{\omegai}{\omega_i}
\newcommand{\Omegas}{\Omega^{(soil)}}
\newcommand{\Omegasn}{\Omega^{(snow)}}
\newcommand{\Omegac}{\Omega^{(cryoconite)}}
\newcommand{\Omegaw}{\Omega^{(bulk\ water)}}
\newcommand{\omegaw}{\omega^{(bulk\ water)}}

\newcommand{\Omegal}{\Omega_l}
\newcommand{\Omegag}{\Omega_g}
\newcommand{\Omegar}{\Omega_r}

\newcommand{\Omegai}{\Omega_i}
\newcommand{\bsa}{\begin{subequations}}
\newcommand{\esa}{\end{subequations}}
\newcommand{\ccup}{\;\overline{\cup}\;}
\usepackage{soul}

\newcommand{\ba}{\begin{eqnarray}}
\newcommand{\ea}{\end{eqnarray}}
\newcommand{\bsub}{\begin{subequations}}
\newcommand{\esub}{\end{subequations}}
\newcommand{\bas}{\begin{eqnarray*}}
\newcommand{\eas}{\end{eqnarray*}}
\newcommand{\R}{\mathbb{R}}
\newcommand{\ave}[1]{\left\langle#1\right\rangle}

\newcommand{\nmat}{N_{MAT}}
\newcommand{\nwmat}{N_{WMAT}}
\newtheorem{example}{Example}
\newtheorem{algorithm}{Algorithm}
\newtheorem{deff}{Definition}

\newtheorem{remark}{Remark}
\newtheorem{assum}{Assumption}

\newcommand{\mpunit}[1]{\mathrm{[#1]}}
\newcommand{\mpcitee}[2]{{\cite{#1}(#2)}}

\newcommand{\pcite}[1]{{\cite{#1}}}

\newcommand{\myskip}[1]{}

\newcommand{\cC}{\mathcal{C}}
\newcommand{\hH}{\mathcal{H}}

\newcommand{\eps}[1]{\varepsilon}

\newcommand{\yY}{\mathcal{Y}}

\newcommand{\tboil}{\theta^*}
\begin{document}

\title
{Heat conduction with phase change \\
in soils with macro-pores, snow, and cryoconite. \\Part I: unified model derivation and examples\thanks{ This research was partially supported by the grants NSF DMS-1912938 ``Modeling with Constraints and Phase Transitions in Porous Media'' and by Joel Davis faculty scholar position, PI: Malgorzata Peszynska}.}

\author{Malgorzata Peszynska, \\Praveeni Mathangadeera, Madison Phelps, \\
Forrest Felsch, Noah Unger-Schulz}

%% MPESZ ORCID 0000-0001-5013-7943
%% PRAVEENI ORCID 0009-0006-1664-2288
%% PHELPS ORCID 
%% FELSCH 0009-0003-9286-2862
%% UNGER-SCHULZ 0009-0003-0212-0313

\maketitle
%%%%%%%%%%%%%%%%%%%%%%%%%%%%%
\begin{abstract}
In this paper we extend models of thermal conduction with phase transition from micro- to macro-scale. Such models were previously developed for soils in permafrost regions from pore to Darcy scale, and the Darcy scale models compare well to empirical relationships.   The new general model blends soil model with bulk water model and thus works well for the soils with macro-pores; it also applies to new context including modeling thermal conduction in the snow and in cryoconite, and it is consistent with rigorous thermodynamics derivations as well as with practical models from the literature. From mathematical point of view, the general model relies on carefully defined relationships between temperature, phase fraction and internal energy,  which we show are invertible in a properly defined framework of multivalued graphs. We also  discuss and test practical models for average heat conductivity.

Our framework allows to create monolithic numerical models useful for modeling of coupled soil and snow models as well as cryoconite.  We motivate and illustrate the results with practical examples and computations.  

\end{abstract}
\noindent
{\bf Declarations}
\\
{\bf Availability of data and material}
\\
Data will be available upon request.
\\
{\bf Competing interests}
\\
The authors declare no competing interests.
\\
{\bf Funding}
\\
This research was partially supported by the grants NSF DMS-1912938 ``Modeling with Constraints and Phase Transitions in Porous Media'', NSF DMS-2309682 ``Computational mathematics of Arctic processes'' and by Joel Davis faculty scholar position, PI: Malgorzata Peszynska, and by Oregon State University URSA program.
\\
{\bf Authors' contributions}
\\
The research was conducted by the authors, with the effort proportional to the order in which their names are listed. Peszynska designed and led the research and contributed to the majority of the writing. Mathangadeera, Phelps, Felsch and Unger-Schulz  participated in the research and in the writing. 
\\
{\bf Acknowledgments}
\\
The authors wish to thank colleagues who were involved in the discussions as well as in earlier steps of this research: Naren Vohra, Lisa Bigler, Zachary Hilliard, Corbin Savich, and Hannah Dempsey. 

\section{Introduction}

In this paper we give new perspectives on modeling heat conduction with phase transitions in complex composite media at micro and macro-scale.  Our work so far has been primarily motivated by the phenomena in the Arctic soils in the areas of permafrost. These respond to the atmospheric controls through the boundary conditions and depend on the presence of snow. However, there are few comprehensive models for both: soils and snow. Derivation of such a model from micro- to macro-scale serves as the motivation for this paper.

We extend our work in \cite{BPV,PVB} on air-free soils to a larger context including soils with macro-pores and snow; we also consider cryoconite, an important partially organic substance covering portions of ice present in the Arctic and high alpine environments.    All these applications involve heat conduction in heterogeneous materials, and the properties of these materials depend on the scale at which they are considered and the phase of the components from which they are made.

\medskip
 The heat conduction model 
for the conservation of energy combined with Fourier law reads
\bsa
\label{eq:heat}
\ba
\partial_t w - \nabla \cdot (k \nabla \theta)&=&0, \; (x,t) \in \Omega \times(0,T]
\\
w(x,0)&=&w_{init}(x), \; x \in \Omega; \; \theta (x,t)=\theta_D(x,t), \; x \in \partial \Omega; t \in (0,T],
\ea
\esa
and is solved for temperature $\theta$ and enthalpy $w$. The model has to be closed with a relationship between $w$ and $\theta$. In a single material in one phase we have $w =  \cC(\theta)$ where $\cC(\cdot)$ is a nonlinear capacity term, with the heat capacity $c=\frac{d\cC}{d\theta}$. In multiple materials $c$ involves volume weighted averages $\ave{c}$. For materials undergoing phase transitions such as freezing and thawing at temperature $\tfr=0$, we have $w =  \cC(\theta)+L\chi$
which involves liquid fraction $\chi$, and heat capacity term 
$\cC(\theta)$, a smooth function of $\theta$, and latent heat $L$. Additionally, we require the  heat conductivity $k=k(\theta)$ which is typically nonlinear function of $\theta$. The model also requires initial conditions $w_{init}$ and boundary conditions for $\theta$. 

The most interesting and mathematically challenging part of \eqref{eq:heat} is the constitutive relationship between  $w,\theta$ and $\chi$. This feature has been very well studied, and well-posedness of \eqref{eq:heat} has been analyzed in particular for the well-known Stefan problem where $\chi \in \hH(\theta)$ is modeled with the multivalued Heaviside graph $\hH$ \cite{Vis2007,Showalter:Stefan}; here, the symbol $\in$ is in place of an equality used for functions. With this feature, $\theta(x,t)$ is continuous, but $w(x,t)$ features a jump across the free boundary where $\theta(x,t)=0$, thus, typically $w \in L^{\infty}(L^2)$ is the best regularity one can expect; see \mpcitee{Vis2007}{Thm.~3.1}. These are well seen in numerical illustrations, e.g., those with conservative fully implicit finite volume scheme in \cite{BPV,VP-Tp}, where $w$ is a primary unknown. In turn, $\theta$ can only be a primary unknown in models where some model approximations of $\theta \to w$ are used. 

In composite soils with many phases and components, the constitutive properties for  $\cC(\cdot),\chi, k(\theta)$ in \eqref{eq:heat} depend pointwise on those of these phases and components. If tracking them pointwise is not practical, one can derive effective constitutive properties $\chieff,\cCeff,\keff$, e.g.,  through  homogenization \cite{BLP,Horn97}. For problems with phase change, the papers \cite{Damlamian81,Vis2007} were the first to prove rigorous results for the effective models. At the same time, effective models with empirical properties $\chiemp,\cCemp,\kemp$ at Darcy scale have been routinely used in reservoir simulation literature \cite{lake}, geotechnics \cite{bookFGE}, and in hydrology of Arctic regions \cite{Osterkamp1987,JafarovMarchenkoRomanovsky12, LingZhang}.

In our recent work \cite{PVB,PHV} we applied the theory in \cite{Vis2007} to modeling air-free soils from pore to Darcy scale and obtained $\chieff,\cCeff,\keff$ which compared well to empirically calibrated relationships $\chiemp,\cCemp,\kemp$. However, we only considered data for which the limit $\wt{\chieff}$ of $\chieff$ was continuous and a.e. differentiable, thus  $\theta$ and $w$ were both continuous. For such situations, the approximation algorithm in \cite{BPV,VP-Tp} could use either $w$ or $\theta$ as a primary unknown; in fact, in \cite{PM24} we were able to compare fully implicit and sequential  approaches using either $\theta$ or $w$.  

{\bf Objectives.} In this paper, we extend the models from \cite{PVB} to  the realistic case where soil includes (a) pockets of air and (b) macro-pores. While (a) requires only a minor revision of the effective properties, (b) gives $\chieff$ with a multivalued limit $\wt{\chieff}$. 
In consequence, the numerical model can only use $w$ as a primary unknown and  requires a delicately tuned local nonlinear solver to resolve the relationships $(\theta,\chi) \leftrightarrow w$.   On the other hand, the new generalized model applies also immediately to two new contexts: (c) modeling thermal processes in pure snow as well as in (d) in cryoconite (``dirty snow'') for which we develop and connect micro to macro-scale models. The new model also outlines the possible effective $\keff$ with some practical approximations $\wt{\keff}$.

This paper serves as part I of the discussion of the generalized model setting the stage for part II where we study the numerical approximation and test the solvers. 

Furthermore, we need data to test the effective models and their approximations and compare them to the empirical models. However, unlike other materials that can be tested and imaged in laboratory conditions, the thawing/freezing soils, snow, and cryoconite require special conditions for testing and experiments, and therefore empirical data especially at the micro-scale is scarce. Therefore, we create synthetic data with properties consistent, e.g., with \cite{calmels,Rooney22}.  

While we focus here only on the heat conduction models, more general coupled models involve the flow and mechanical deformations; we refer to \cite{Garayshin19,PHV,HEP} for these more general contexts where our new general model based on first principles will eventually be coupled with other phenomena.    
Finally, there are other applications where this research might be useful including biological tissue cryo-preservation, food preservation, as well as studies of extraterrestrial environments.

{\bf Outline.} In Section~\ref{sec:notation} we define the necessary notation. In Section~\ref{sec:materials} we describe the materials of interest to this paper: soils, snow and cryoconite. In Section~\ref{sec:models} we discuss the challenges of modeling multiple phases and in Section~\ref{sec:newmodels} we define new models. In Section~\ref{sec:examples} we present examples.  

{\bf Notation.} For a set $A$ we denote by $\abs{A}$ its measure or cardinality, by $\bar{A}$ its closure, and by $\partial A$ its boundary. 
For non-overlapping subdomains $A,B \subset \R^d$, if $A \cap B=\emptyset$, but their interface $\partial_{AB} = \partial A \cap \partial B  \neq \emptyset$, instead of $S=A \cup B \cup \partial_{AB}$ we shall abbreviate $\Omega = A \ccup B$. For a subset $A \subset S$ we denote by $\onechar_A$ the characteristic function  as $\onechar_A(x)=1, x \in A$ and equal 0 otherwise. 

%%%%
\section{Composite materials: multiple components and phases}
\label{sec:notation}

We consider heat conduction \eqref{eq:heat} in a spatial domain, an open bounded set $\Omega \subset \R^d, d\geq 1$.  The boundary of $\Omega$ is denoted by $\partial \Omega$ which we assume to be entirely of Dirichlet type.
We will use $x \in \Omega$ to denote the spatial variable at the macro-scale. 

We will also work in a micro-domain $y \in \omega(x)$ where $\omega(x)$ is an open simply connected region located at $x$, with small characteristic length: $diam(\omega) \ll diam(\Omega)$, but large enough to be an REV (Representative Elementary Volume) \cite{BearCheng10}. A convenient conceptual model is that the macro-scale region is a union of the REVs with $\overline{\Omega}=\bigcup_x \overline{\omega}(x)$, and that each $\omega(x)$ is a cube $[0,1]^3$ scaled by some $0 <\epsilon\ll 1$ and translated to $x$; this is the view taken in homogenization theory \cite{Tartar,Horn97}. 

The model \eqref{eq:heat}  requires data for $\cC(\theta)$ as well as for the conductivities $k(\theta)$; the variables and data are listed in Table~\ref{tab:variables}.  

%%%
\begin{table}[htp]
\begin{tabular}{|l|l|l|}
\hline
\hline
Variable & Description & SI Unit
\\
\hline
\hline
$w$ & Enthalpy/energy per unit volume & $\mpunit{J / m^3}$
\\
\hline
$\theta$ & Temperature& $\mpunit{^\circ C}$ 
\\
\hline
$\chi$ & Phase fraction & $\mpunit{-}$
\\
\hline
\hline
Data & Description & SI Unit
\\
\hline
$\cbar,c$&heat capacity (per unit mass or volume)& $
\mpunit{J /(kg \cdot K)}$, $
\mpunit{J / (m^3 \cdot \degK)}$
\\
$\cC(\theta)$& smooth part of $w$ per unit volume & $\mpunit{J / m^3}$ 
\\
$L$& latent heat at phase change temperature $\tfr$& $\mpunit{J / m^3}$ 
\\
$k$& heat conductivity & $\mpunit{W /(m \cdot \degK)}$
\\
\hline
$x,y$&macro- and micro spatial scale variables&1$~\mpunit{m}$,$1$-$10^4~\mpunit{\mu m}$.
\\
\hline
\end{tabular}
\caption{Variables and data in heat conduction models. We adopt the convention that ``phase fraction'' in ice-liquid phase change models means liquid phase fraction, and in liquid-vapor models it means the vapor  phase fraction.}
\label{tab:variables}
\end{table}

In this paper we make a distinction between materials $m=1, \ldots \nmat$ denoted by superscript $(m)$, phases  $p=l,i,g$ (liquid, ice, and gas) denoted by subscript $p$, and  components $C=R,W,A$  (rock, water, air) denoted by superscript $C$.
In particular, we consider volumetric fractions $\etam$ of materials, phase  volume fractions $\eta_p^{(m)}$, mobile phase saturations $S_p$, and mass fractions $\mymu_p^C$. This well known notation is summarized in  Table~\ref{tab:mixture}. 

The thermal model \eqref{eq:heat} at a micro-scale can use $c,k$ dependent on the properties of the components in the individual phases which can be recognized pointwise: modeling how a component changes phase is the crux of the problem. At a macro-scale, the data for $c,k$ are appropriate weighted averages of micro-scale properties, and empirical models typically postulate some selected weighting.  In this section we set up the background for micro- and macro-scale models.  

%%%%%
\subsection{Single material} 
For a single material in one phase, we have 
\ba
\rho = const,\; \cbar= const,\; k = const; \;\; \cC(\theta)=\cbar \rho \theta = c\theta. 
\ea
Here $\cbar$ is heat capacity, $\rho$ density, which we assume constant, but recognize that  in some scenarios it is necessary to account for their dependence on the temperature. For example, in ice, \mpcitee{CRC}{6-12,16} $k_i=2.16$ at $\theta=0$ but $k_i=2.26$ at $\theta=-10$. 

%%%
\subsection{Multiple materials} 
\label{sec:multim}

\begin{table}[htbp]
\footnotesize
\centering
\begin{tabular}{|c|l|}
\hline
Notation & {Description}
\\
\hline
$(m)$ & {Superscript for subdomain occupied by material numbered by $m$}
\\
$\omega$& A REV domain: an open bounded domain 
\\
$\omegam$, $\Omegam$& {Subdomain occupied by material $m$}
\\
$\eta^{(m)}$& {Volume fraction of material $m$ in REV $\omega$ : $\eta^{(m)}=\frac{\abs{\omega^{(m)}}}{\abs{\omega}}$ }
\\
\hline
&Notation for multiple phases 
\\
\hline
$p$&Subscript for phase properties of phase $p$
\\
$\omega_p$ & {Subdomain $\omega_p \subset \omega$ of the REV $\omega$ occupied by phase $p$}
\\
$\eta_p=\frac{
\abs{\omega_p}}{\abs{\omega}}$& Volume fraction of phase $p$ in $\omega$ ; \; $\sum_p \eta_p=1$ 
\\
\hline
&Notation for multiple phases in multiple materials 
\\
\hline
$\omega^{(m)}_p$ & {Subdomain $\omega^{(m)}_p \subset \omega$ occupied by phase $p$ of material  $m$ in the REV $\omega$ }
\\
$\eta_p^{(m)}=\frac{
\abs{\omega^{(m)}_p}}{\abs{\omega}}$& Volume fraction of phase $p$ of material $m$ in $\omega$ 
\\
\hline
&Notation for multiple components
\\
\hline
$\mymu_p^C$&Mass fraction of of component $C$ in phase $p$ 
\\
\hline
\end{tabular}
\caption{Notation for multiple materials $m=1,\ldots \nmat$, phases, and components.}
\label{tab:mixture}
\end{table}
%%%%%%%%%%%%%%

Consider a composite (heterogeneous) material made of $\nmat$ homogeneous materials, in which the position of materials  does not change in time, e.g., with natural or man-made geometry forming a layered or a periodic medium. 

In a composite material a property $\myp=\myp(x)$ such as $\rho,c,k$ can be described piecewise $\myp(x) = \sum_m \myp^{(m)} \onechar_{\Omega^{(m)}}(x),
x \in \Omega$, and we require  $\rho^{(m)}, k^{(m)}, c^{(m)}$. 
If the scale at which $\myp$ varies is small, then the coefficients  $c(x,y),k(x,y)$ depend on the large scale $x \in \Omega$ and small scale $y \in \omega(x)$ and are piecewise  in $\omegam$: $\myp(x,y) = \sum_m \myp^{(m)} \onechar_{\omega^{(m)}(x)}(y)$. At the micro- or macro-scale, if   $\eta^{(m)}(x)=\frac{\abs{\omega^{(m)}}(x)}{\abs{\omega(x)}} \approx const$, we denote these variations by
\ba
\label{eq:mypm}
\myp = \left( \eta^{(m)},\myp^{(m)} \right)_{m=1}^{\nmat} = 
\left[ \begin{array}{cc}
\eta^{(1)} & \myp^{(1)}
\\
\eta^{(2)} & \myp^{(2)}
\\
... & ... 
\\
\eta^{(\nmat)} & \myp^{(\nmat)}
\end{array}
\right].
\ea
%%%

%%%%
\subsubsection{Material volumetric averages}
\label{sec:averages}
For composite materials, it is useful to consider average properties along the depth or over the REV
\bas
\label{eq:ave}
\ave{\myp}(x)= \tfrac{1}{\abs{\omega(x)}} \int_{\omega(x)}\myp(x;y)dy.
\eas
%%%
For multiple materials in \eqref{eq:mypm} we have arithmetic
and geometric averages, respectively
\begin{subequations}
\label{eq:aveag}
\ba
\label{eq:avea}
\myp 
&\stackrel{[A]}{\to} &\ave{\myp}^{[A]}=\sum_{m=1}^{\nmat} \eta^{(m)}\myp^{(m)};
\\
\label{eq:aveg}
\myp 
&\stackrel{[G]}{\to} &
\ave{\myp}^{[G]}
=\Pi_{m=1}^{\nmat} (\myp^{(m)})^{\eta^{(m)}}.
\ea
\end{subequations}
One also considers harmonic averages $\ave{\myp}^{[H]}= (\ave{1/\myp}^{[A]})^{-1} $ as well as upscaling $\myp \stackrel{[UP]}{\to} \ave{\myp}^{[UP]}$ discussed in Section~\ref{sec:upscale}. 

%%%
\subsection{Multiple phases of a single component or material (substance)} 

Materials can exist in multiple phases. The phases have distinct physical properties such as density $\rho_p$ and thermal properties $c_p$ and $k_p$. The phases do not mix (are immiscible) and the phase interfaces are visible at some fine enough scale; such is the case of oil (hydrocarbon) and water in normal conditions; however, at high pressures or temperatures these phases may become miscible and may mix.   The partition of a material between phases depends on the pressure, temperature, and other factors. 

Consider a single material $m=1$ made of a single component in multiple phases so that a property $
\myp(x,t) = \sum_p \myp_p \onechar_{\Omega_p (x,t)}.
$, but the position and phase volume fractions may change in $(x,t)$. Similar properties are at a micro-scale  $
\myp(x,y,t) = \sum_p \myp_p \onechar_{\omega_p (x,y,t)}$. We  discuss the formal definitions of the subdomains below. We also bring up the practical challenges on modeling involving pointwise temperatures at the micro-scale and equilibrium with average temperatures. 

%%%
\subsubsection{Bulk water domains $\omegaw,\Omegaw$}
\label{sec:omegaw}
At atmospheric pressure, water $W$ in a macro-scale domain $\Omegaw$ can (in principle) exist in the ice, liquid, or gas (vapor) phases depending on the temperature $\theta$ with respect to the freezing/thawing temperature $\tfr$ and boiling/condensation temperature $\tboil$. Similar phase partition is at micro-scale in $\omegaw$. The phase interfaces at time $t$ such as $\gamma_{il}(t)=\{y: \theta(y,t)=\tfr\}$ and $\gamma_{lg}(t)=\{y: \theta(y,t)=\tboil\}$ between the ice in $\omegai(t) \ni y: \theta(y,t)<\tfr$, liquid $\omegal(t) \ni y: \tfr< \theta(y,t) < \tboil $ and vapor $\omegag(t) \ni y: \theta(y,t)>\tboil$, respectively, are determined depending on $\theta(y,t)$. One can imagine coexistence of all three phases, e.g., in a pot of boiling water where we toss in a few ice cubes, with the temperature highly varying until it reaches some equilibrium. 

\begin{figure}[htp]
\centering
\includegraphics[height=4cm]{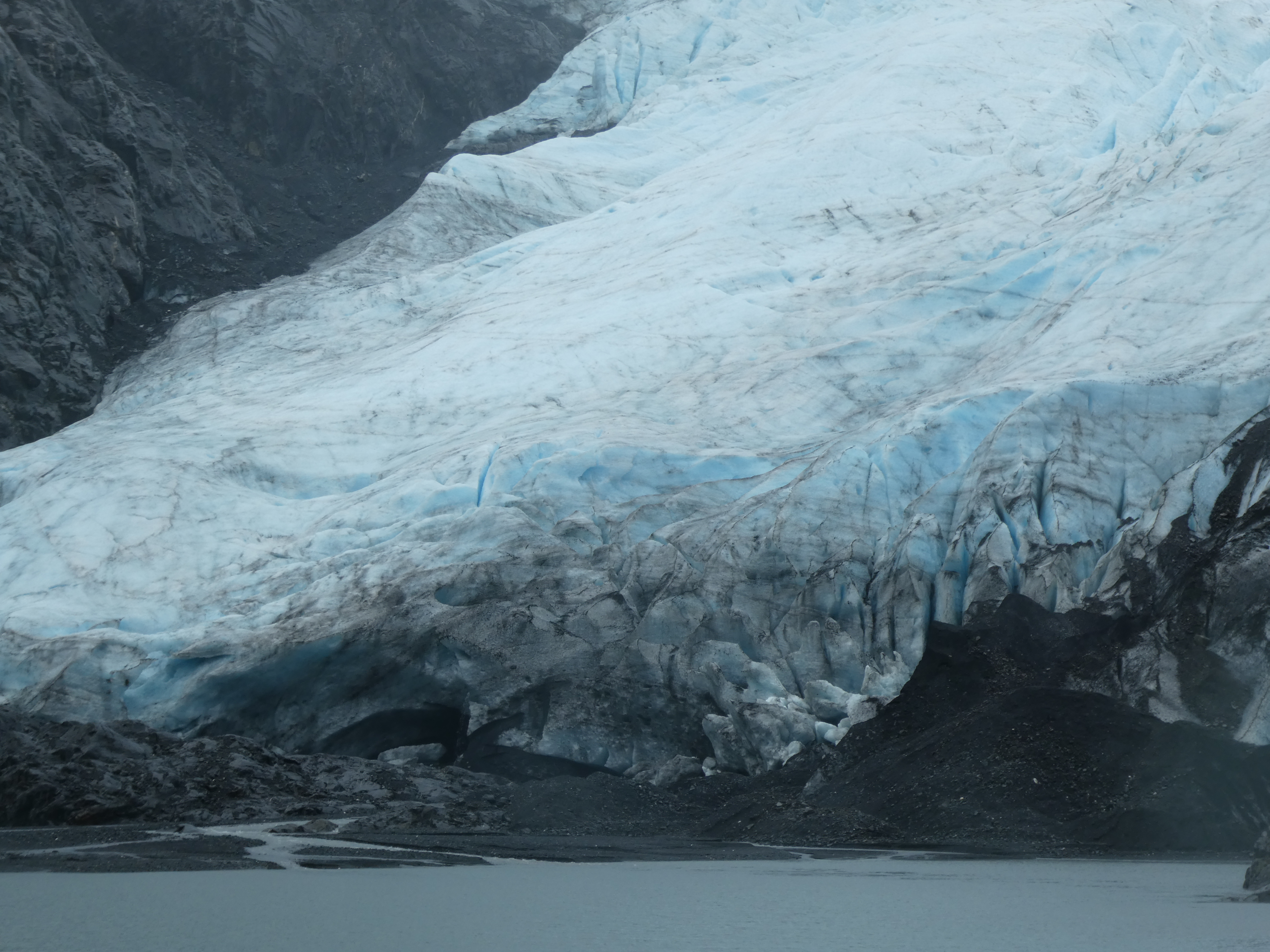}
\includegraphics[height=4cm]{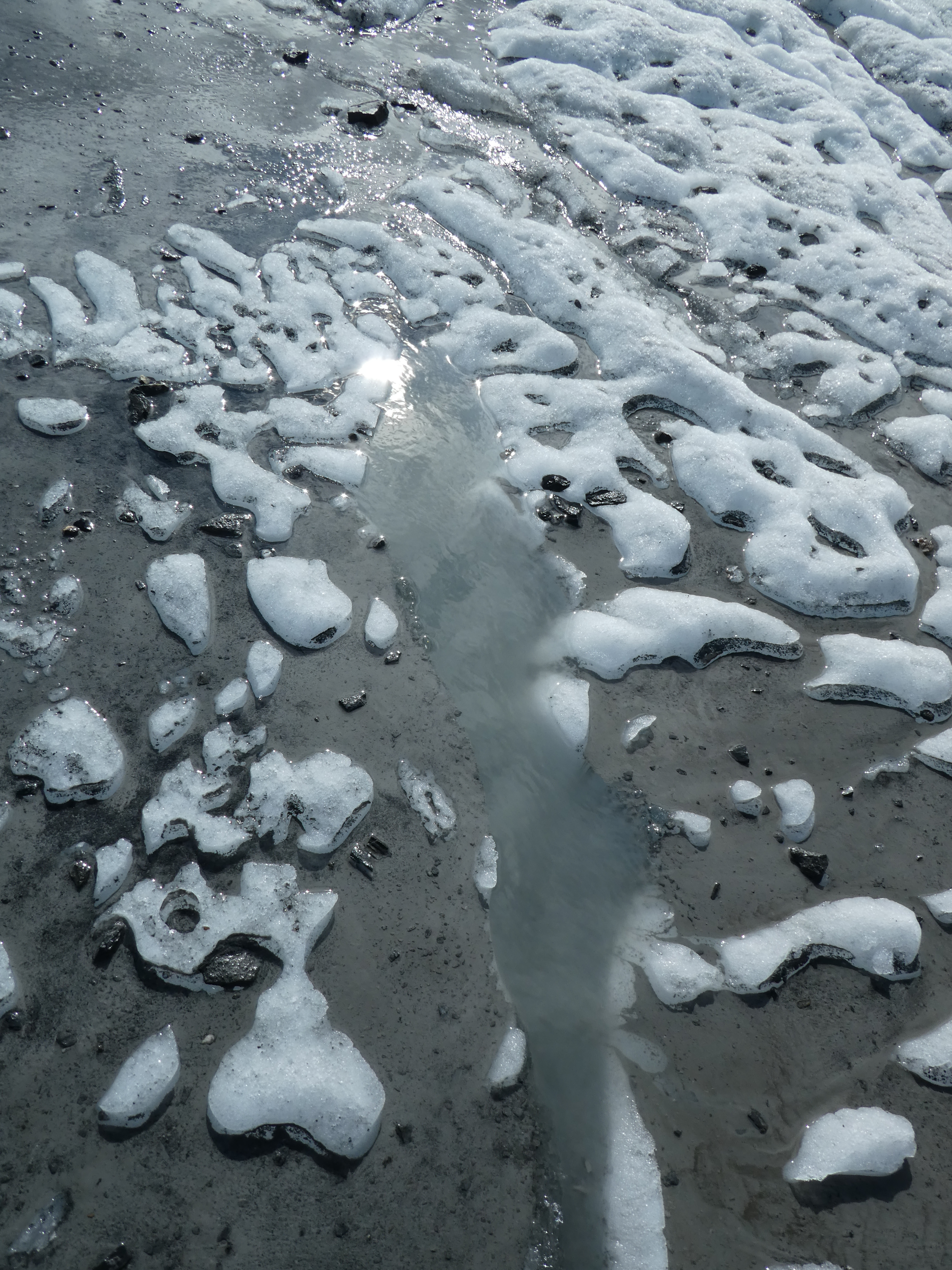}
\includegraphics[height=4cm]{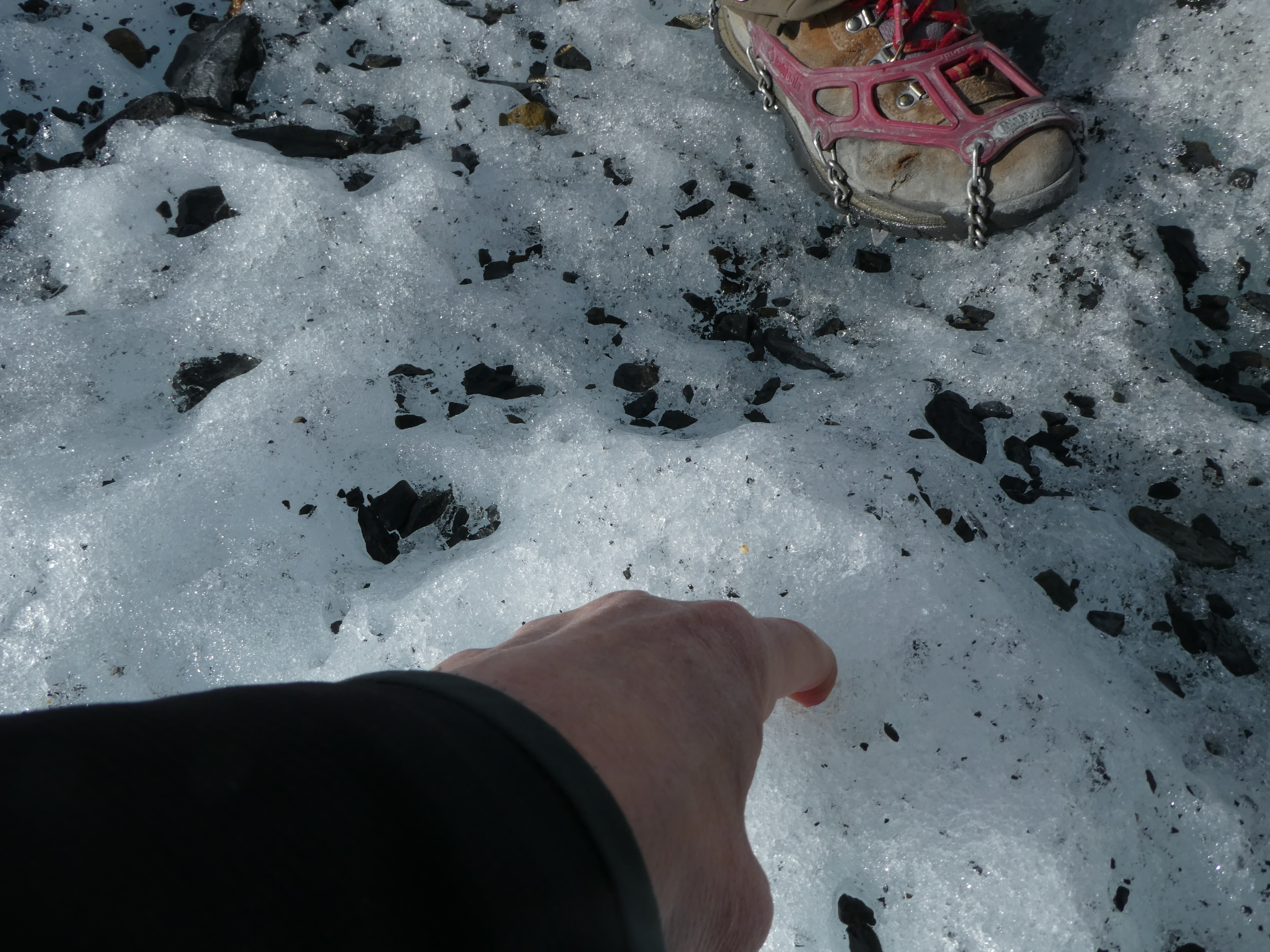}
\caption{Multiple scales: Domain $\Omega$ with clearly delineated subdomains $\Omegai$, $\Omegal$, $\Omegar$ at a very large scale (left, $100~\mpunit{m}$, Portage glacier), large scale (middle, $1~\mpunit{m}$ silty approach to Matanuska glacier). Right: dirty snow/cryoconite image from Matanuska glacier, micro scale is that of ice crystals.  (Photo credits: MP, September 2024, Alaska.)}
\label{fig:scales}
\end{figure}

In  a realistic natural environment however, the range of $\theta(y,t), y \in \omega(x)$ is rarely large enough to include all $i,l,g$ phases of $W$, but it is easier to accept that there is an approximate equilibrium of the phase distribution with the REV average $\ave{\theta}(x)$.  At a very large macro-scale this is also rare and the phase distributions of $\Omegai,\Omegal,\Omegag$ are not in equilibrium with any average temperature, as illustrated, e.g., in Figure~\ref{fig:scales}, with liquid, ice, and gaseous regions undergoing an evolution  towards equilibrium which is most slow at the largest scales.  

\begin{remark}[Scales and equilibrium]
\label{rem:EQscales}
In this paper we aim to work with macro-scale models for which it is not practical to resolve phase interfaces at the micro-scale but at the same time it is reasonable to assume equilibrium of average phase fractions with the average temperature, and  phase averaged properties of materials.  
\end{remark}

%%%
\subsection{Notation for graphs} 
\label{sec:graphs} 

We define the notion of {\em relation} or {\em graph}
as a generalization of function.
We adopt notation and nomenclature from \mpcitee{PShin21}{Section 2}, but  refer to the comprehensive details on the general abstract Hilbert space setting, working with PDEs with graphs and monotone multivalued operators, e.g., in \cite{Brezis73,Show-monotone,PSW}. This notation is fundamental to our ability to work with a unified model.

A real-valued graph is a subset $G \subset \R \times \R$, a notion useful in describing phase transitions. For graph $G$, its domain and inverse are defined as  $\mathrm{domain}(G)=\{a: {\exists b:} (a,b) \in G\}\subset \R$, $G^{-1}=\{(b,a): (a,b)\in G\}$. 
The graph $G$ is monotone if $\forall (a_1,b_1), (a_2,b_2) \in G$ we have $(b_2-b_1)(a_2-a_1)\geq 0$.  $G$ is
maximal monotone if $G$ is monotone and $I+G$ is onto $\mathbb R$, that is, for every $c \in \mathbb R$, there exists $(a,b) \in G$ such that $a+b = c$. The graph $G$ is multivalued (also called ``set-valued'') if $\exists a \in \mathrm{domain}(G):$ the cardinality of $G(a) \equiv \{b:\ (a,b) \in G\}$ is more than $1$; this means that $G$ cannot be interpreted  as a ``graph'' of some function. If $G$ is a graph and $(a,b) \in G$, we write $b \in G(a)$ to denote some (non-unique) selection $b$ out of the set $G(a)$. Note that a graph $G$ is monotone or maximal monotone if and only if $G^{-1}$ is, respectively, monotone or maximal monotone.

%%%
\begin{example}[Heaviside function and Heavisde graph]

We will use the Heaviside function $H:\R\to \{0,1\}$, $H(0,\infty)=\{1\}; H(-\infty,0]=\{0\}$ which is discontinuous at $0$. 
The multivalued Heaviside graph with domain $\R$ is maximal monotone and $(I+\epsilon \mathcal H)^{-1}(x)$ equals $x$ if $x< 0$, equals $0$ if $0 \le x \le \epsilon$, and equals $ x-\epsilon$ if $x > \epsilon$. Also, 
\ba
\label{eq:hH}
\hH = \left[(-\infty,0) \times \{0\}\right] \cup 
\left[
\{0\} \times [0,1] \right] \cup 
\left[(0,\infty) \times \{1\}\right], \;
v \in \hH(u) =
\left\{ \begin{array}{ll}
0, &u<0,
\\
{[}0,1{]}, & u=0,
\\
1, &u<0.
\end{array}
\right.
\ea
However, $\hH^{-1}$ with domain $[0,1]$ is maximal monotone.
\end{example}

%%%%%%
\subsection{Graphs for phase transitions for single component in a composite material} 
As we described in Section~\ref{sec:omegaw}, the phase transition in ice-liquid domains $\Omegaw$ or $\omegaw$ occurs at a single temperature $\theta=\tfr$ so the liquid phase fraction $\chi \in\hH(\theta-\tfr)$ is multivalued.  

However, in soils $\Omegas$, the freezing temperatures are distributed depending on the characteristic distribution of pore sizes in each $\omegam(x)$ in each REV $\omega(x)$, and one can consider several $\tfrm(x)$ for each $x$. Therefore, with a large number of pore sizes in each REV, the liquid phase fraction $\chi=\chi(\theta)$ can be  a smooth function reflecting this property, and empirical models $\chiemp$ called Soil Freezing Curves (SFC) are calibrated based on this (equilibrium) assumption. 

In this paper we aim to build a general model to account for different types of phase transitions including in $\Omegaw$ and in $\Omegas$ with a single freezing temperature and a distributed set of temperatures.   The model will be given in detail in Section~\ref{sec:models}, but we set the stage now by defining notation.
The general model involves parameters $\tfr,\chis$ and a function $\Upsilon$. These are selected based on the following Assumption, with the definition of a graph $\gG$ generalizing $\hH$ to follow.   

%%%%
\begin{assum}
\label{ass:upsilon}
Let 
$
\tfr\leq 0$ and $0 \leq \chis \leq 1.
$
be given. Also,  let $\Upsilon :\R_- \to [0,\chis]$ be a smooth monotone {nonde}creasing function 
 with 
$ \lim_{\vartheta \to 0^-}\Upsilon(\vartheta) =\chis$, and $\lim_{\vartheta \to -\infty}\Upsilon(\vartheta) =0$.
If $\chis>0$ we will assume that $\Upsilon$ is strictly increasing.
\end{assum} 

\begin{figure}[htp]
\centering
\includegraphics[height=4cm]{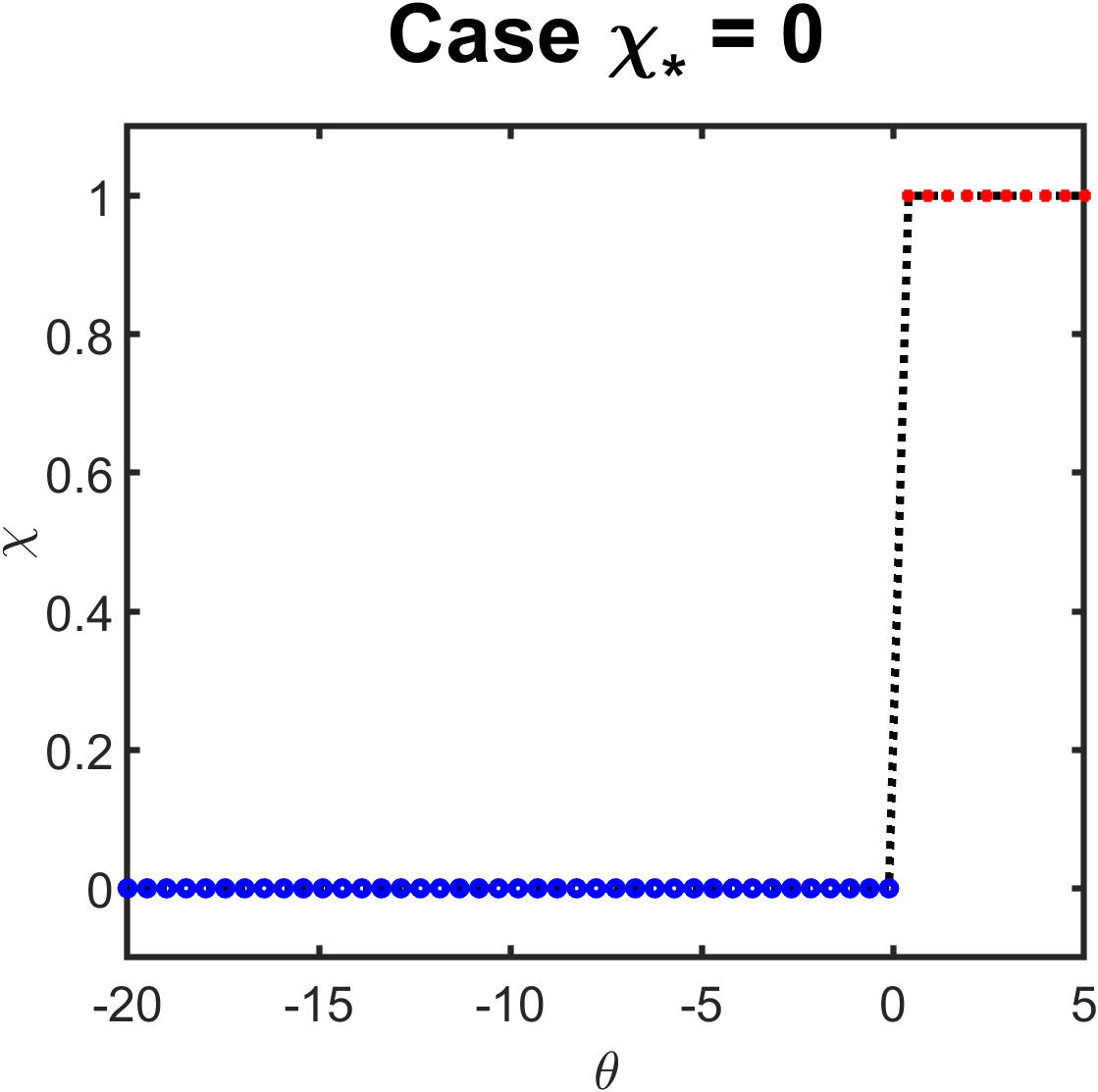}
\includegraphics[height=4cm]{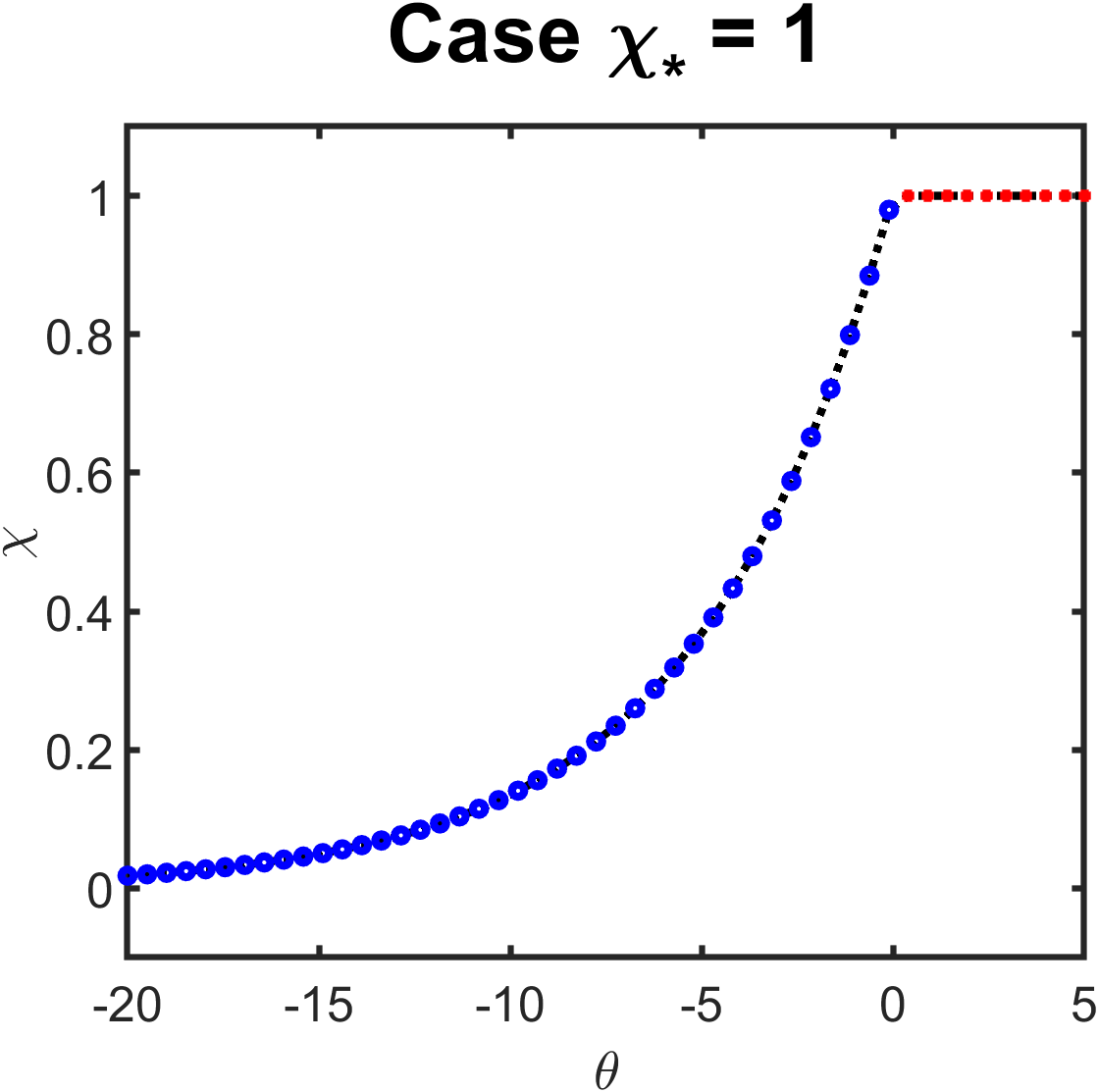}
\includegraphics[height=4cm]{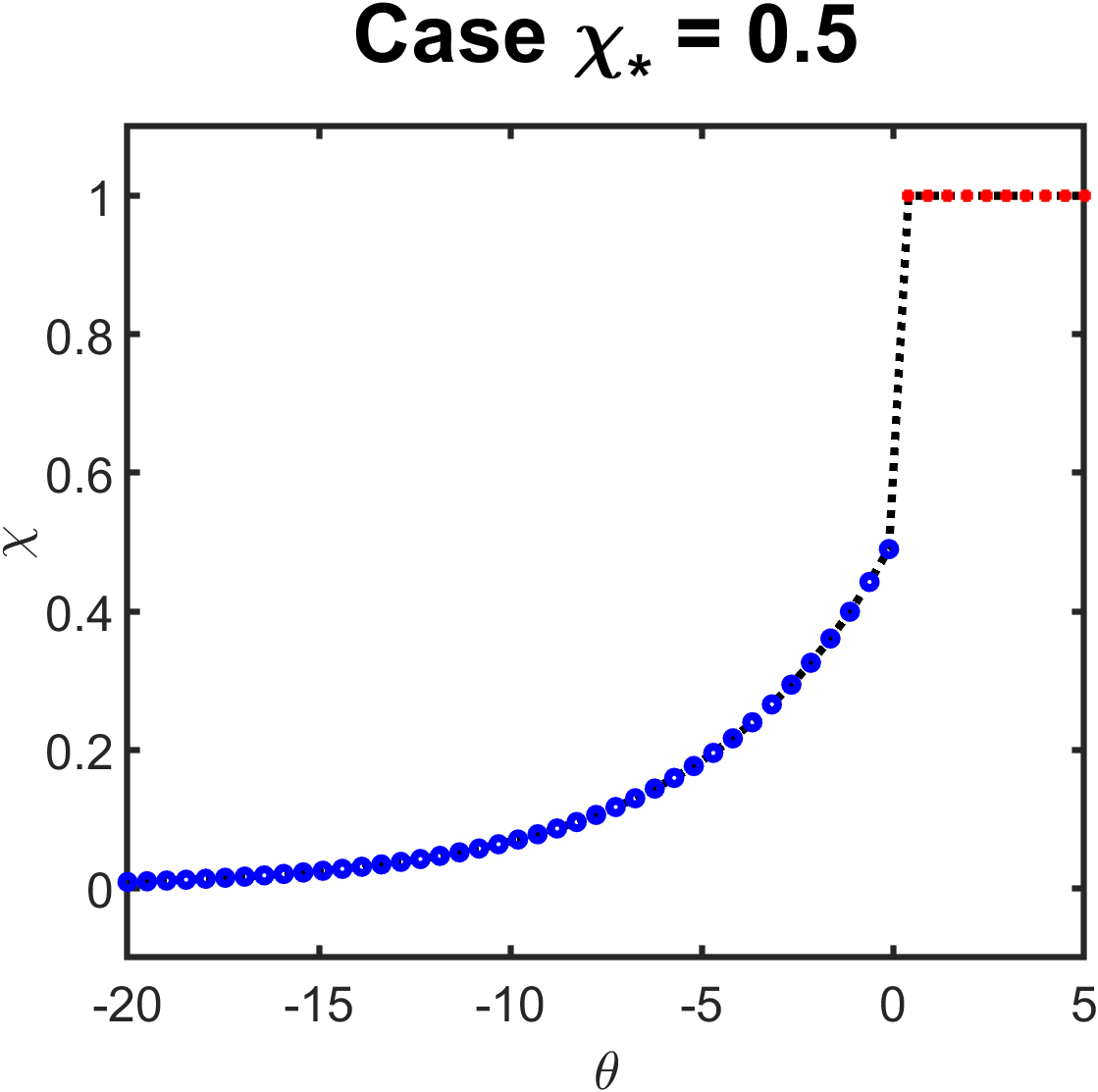}
\caption{Plot of $(\theta,\chi) \in \gG[\tfr,\chis,\Upsilon]$ when $\tfr=0$ satisfying \rass{ass:upsilon} for the three cases in Example~\ref{ex:chizero} ($\chis=0$), Example~\ref{ex:chisone} ($\chis=1$), and Example~\ref{ex:chiPs} ($\chis \in (0,1)$), respectively. For $\chis>0$ we choose $\Upsilon(v)=\chis e^{b v}$ with $b=0.2$.}
\label{fig:graph}
\end{figure}

%%%
\begin{deff}
\label{def:chi}
Let $\Upsilon$ satisfy \rass{ass:upsilon}. We define the graph $\gG\subset \R \times [0,1]$ as follows
\ba
\label{eq:chis}
(\theta,\chi) \in \gG =
\left\{ \begin{array}{ll}
\Upsilon(\theta-\tfr), &\theta<\tfr,
\\
{[}\chis,1{]}, & \theta = \tfr,
\\
1, &\theta>\tfr.
\end{array}
\right.
\ea
Since $\gG$ depends on $\tfr,\chis,\Upsilon$, we denote this by 
$\chi \in \gG[\tfr,\chis,\Upsilon](\theta)$ or write $\chi \in \gG(\theta)$.
\end{deff}
In Definition~\ref{def:chi} we have three special cases $\chis=0,\chis \in (0,1), \chis=1$, which we illustrate in Figure~\ref{fig:graph}.  
 Regardless, 
$\gG$ is not smooth (thus not a smooth manifold within $\R^2$), but it can be parametrized by a single variable $\psi$; for example, on $\hH=\gG[0,0,0]$ the variable $\psi$ is set first to $\theta$, then to $\chi$ at $\theta=\tfr$, and then to $1+\theta$.   

%%%%%%

%%%
\begin{example}[Case $\chis=0$.]
\label{ex:chizero}
Now \rass{ass:upsilon} forces $\Upsilon\equiv 0$, thus actually $\gG(\theta)$ is a shifted $\hH(\theta-\tfr)$, i.e., $\chi \in \hH(\theta-\tfr)$. This case models a typical relationship for liquid fraction $\chi$ at ice-liquid phase transition, or gas fraction in a liquid-gas phase transition in bulk material containing only water to be revisited in Section~\ref{sec:omegawil}. The graph 
\ba
\label{eq:chizero}
\gG[\tfr,0,0] =  
\{ (\theta,\chi): \theta \in (-\infty,\tfr], \chi=0 \}
\cup \{\tfr\} \times (0,1)
\cup (\tfr,\infty) \times \{1\} 
\ea
and its inverse $\gG^{-1}$ are maximal monotone and multi-valued.
\end{example}

\begin{example}[Case $\chis=1$.]
\label{ex:chisone}
When $\chis=1$, $\theta \to \chi(\theta)$ is a continuous function with graph 
\ba
\label{eq:chiP}
\gG[\tfr,1,\Upsilon] =  
\{ (\theta,\chi): \theta \in (-\infty,\tfr], \chi=\Upsilon(\theta-\tfr) \}
\cup [\tfr,\infty) \times \{1\} . 
\ea
Since $\Upsilon$ is  monotone  strictly increasing, the restriction $\chi^{-1}:(0,1)$ is a function but $\chi^{-1}$ is multivalued at $\chi=1$. This case is important for modeling thawing/freezing in soils discussed in Section~\ref{sec:SFC}. 
\end{example}

\begin{example}[Case $\chis\in(0,1)$]
\label{ex:chiPs}
Now $\chi$ and its inverse are multivalued. We have 
\begin{multline}
\label{eq:chiPs}
\gG[\tfr,\chis,\Upsilon] =  
\{ (\theta,\chi): \theta \in (-\infty,\tfr], \chi=\Upsilon(\theta-\tfr) \}
%\\
\cup \{\tfr\} \times [\chis,1] 
\cup (\tfr,\infty) \times \{1\} . 
\end{multline}
On the other hand, if $\chis>0$ and $\Upsilon$ is increasing, then  $\theta \to \chi$ is a function invertible on  $(-\infty,0)$.  
This relationship extends  Example~\ref{ex:chisone} and is well designed to describe soils with large and a variety of small to nano-pores; see Section~\ref{sec:SFC}. 
\end{example} 

%%%%%%%%%%%%%
\subsection{Multiple components within a phase}
Phase such as $p=i$ are made of a single component $W$. However, other phases are made of multiple components, e.g., the atmosphere is in gaseous phase made of different component gases. To distinguish the properties of a regions occupied e.g. by multiple components such as oxygen and water, we need the phase diagram such as in Figure~\ref{fig:porous} describing the possible phase/component partitions as well as thermodynamic data supplementing the description with data on $\mymu_p^C$, usually subject to a maximum constraint, as explained in \cite{lake,Garg08,PSW,PShin21} in a general context of multiple components partitioned between several phases. 

\begin{example}
\label{ex:pvt}
For example, the calculation of heat capacity
$
c(x,t)=\cbar^A_g \rho_g \mymu_g^A + 
\cbar^W_g \rho_g \mymu_g^W
, \; x \in \Omegag.
$
involves the properties of air component (itself a mixture of many gases), and water vapor. 
In turn, 
the water vapor mass fraction cannot exceed a maximum amount $\mymu_g^{W,*}$, and extra water appears as liquid with water fraction $\chi>0$;     
this is described by the relative humidity constraint 
\ba
\label{eq:phasec}
(\chi,\mymu_g^W) \in 
\left\{
\begin{array}{cc}
\mymu_g^W< \mymu_g^{W,*},& \chi=0,\\
\mymu_g^W= \mymu_g^{W,*},&\chi > 0.
\end{array}
\right.
\ea
The maximum amount $\mymu_g^{W,*}$ contained in gas depends on the ambient temperature and pressure. If the amount of $W$ exceeds the maximum $\mymu_g^{W,*}$ or if the temperature decreases, the component $W$ appears as liquid (e.g., rain). 
\end{example}

%%%
\subsection{Modeling heat conduction in composite materials: upscaling to effective properties} 
\label{sec:upscale}

Simplest example of a composite material is a layered  material such as a laminate where averages across its width might be taken. Another good example is soil whose properties are averaged over the rock grains and fluids in the void space, and this can be explained theoretically. 

The mathematical theory of PDEs  in composite materials as well as strategies for finding effective parameters and efficient numerical simulations are very well established; we refer to  \cite{BLP, SanchezPalencia} for general background and to geosciences specific exposition in \cite{Horn97,BearCheng10}, and to recent work on upscaling for coupled processes \cite{Ray2020,BringedalPop,Brun2018UpscalingOT,HEP}. In \cite{PVB} we 
followed \cite{Damlamian81,Vis2007} with focus on thermal models with phase change, and in \cite{PHV} we address
the uncertainty of evolving geometry and construction of surrogate models at the pore-scale following the early attempts in 
\cite{PTISW,BIO2020,PUS}. 

%%%
\subsubsection{Upscaling $c$ and $k$ in a single phase}
\label{sec:upscale-onephase}
For composite materials in a single phase, we have $c= \left( \eta^{(m)},c^{(m)} \right)_m$, and $k= \left( \eta^{(m)},k^{(m)} \right)_m$.   
The macro-scale properties for the linear parabolic problem \eqref{eq:heat} are $\ceff,\keff$ which must be found.   

It is well known \cite{BLP,SanchezPalencia,Horn97} that the appropriate effective conductivity
is $\ceff=\ave{c}^{[A]}$  defined in \eqref{eq:aveag}. For single phase material, this gives the relationship 
\bas
\theta \to w =\cCeff(\theta)=c_{eff} \theta.
\eas
We revisit this formula for multi-phase problems in Section~\ref{sec:ceffm}.

Next we discuss the conductivity $\keff$. 
For homogenization of a two scale problem and upscaling to a coarse problem, \cite{BLP,Horn97,Vis2007} consider dependence of data on $x,y=\frac{x}{\epsilon}$ to be periodic in 
$y\in \yY$ with a unit measure cell $\yY$. For a PDE with an elliptic term involving the conductivity $k(y), y \in \omega(x)$ the effective conductivity $k_{eff}$  
is found via the corrections $(\xi_j)_{j=1}^d$ which solve the auxiliary periodic elliptic problem 
\begin{subequations}
\label{eq:keff}
\ba
\label{eq:keffnaux}
-\nabla \cdot(k(y) (e_j+\nabla_y \xi_j))=0; \;\; \xi_j \text{\ periodic \ on }\yY.
\ea
Then
\label{eq:keffn}
\ba
\label{eq:keffncorr}
k_{eff}=(k_{eff,ij})_{ij};\;\;  k_{eff,ij}=\ave{k(y)(\delta_{ij}+\partial_i \xi_j)},
\ea
\end{subequations}
Simplifications of \eqref{eq:keff} and extensions to nonlinear problems are available, e.g., in upscaling from pore to Darcy scale in \cite{PT13,PTS,TP13}. We revisit this discussion for multiple phases in Section~\ref{sec:keff}.

%%%%%%%%%%%%%%%%%%%%%%%%%%%%%%%%%%%
\section{Materials, phases, and components of interest in this paper}
\label{sec:materials}

Of interest in this paper is heat conduction in composite materials including soil, snow and cryoconite. 
These are heterogeneous materials with the properties $c=c(x),k=k(x)$ in \eqref{eq:heat} which depend on the type of the components and their phases. 
We also need to  consider the scale at which we want to model these phenomena.
We are interested in models at the macro-scale $x \in \Omega$ but we use the micro-scale properties at $y \in \omega(x)$ to motivate and explain our models. 

For example, we have $\Omega^{(soil)}$ and $\Omega^{(snow)}$, with $\cC(\theta)$ and $k=k(\theta)$  defined separately for each domain. Soil may have layers of sand in $\Omega^{(sand)}$ and peat in $\Omega^{(peat)}$, and aggregates of different mineral components characterized by different grain and pore sizes such as in $\Omega^{(clay)}$, $\Omega^{(sand)}$,  $\Omega^{(silt)}$ for which thermal properties vary drastically. 

\begin{remark}
\label{rem:range}
Here we are interested in $\Omegas,\Omegasn,\Omegac$ in close to atmospheric conditions and in the range of temperatures
\ba
\label{eq:thetarange}
\theta \in [-20,10]~\mpunit{\degc}.
\ea
\end{remark}

In Sections~\ref{sec:materialss},  ~\ref{sec:materialsn}, \ref{sec:materialsc}  we describe the thermal properties of soil, snow, and cryoconite, respectively in view of the possible macro-scale description. 
Table~\ref{tab:materials} summarizes these materials.  

Setting aside the mathematical theory, practitioners derive empirical values and nonlinear relationships valid for these materials at the macro-scale.  Others develop micro-scale models. In the subsequent Section~\ref{sec:models} we discuss the effective and empirical thermal properties for multicomponent and multiphase materials and in Section~\ref{sec:newmodels} we focus on the specific materials $\Omegas$, $\Omegasn$, $\Omegac$ discussed here.   
%

%%%%%%%%%%%%%%%%%%%%%%%%%%%
\subsection{Soil domain $\Omegas$: description} 
\label{sec:materialss}

\begin{figure}[htp]
\centering
\includegraphics[height=4cm]{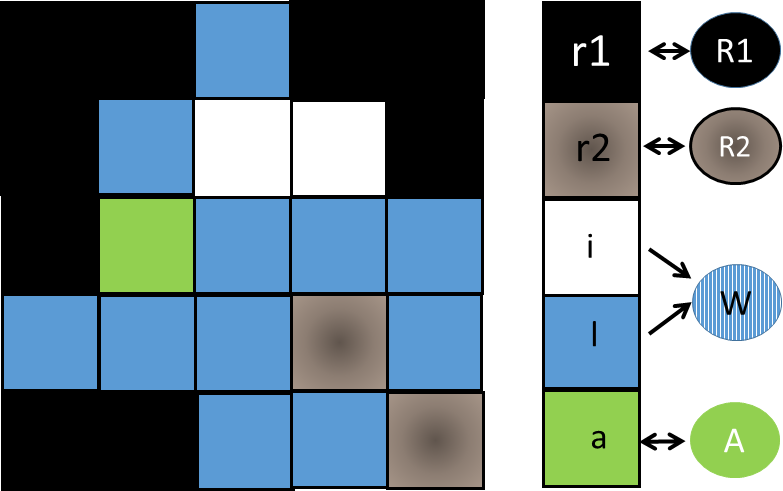}
\;\;\; \;\;\includegraphics[height=4cm]{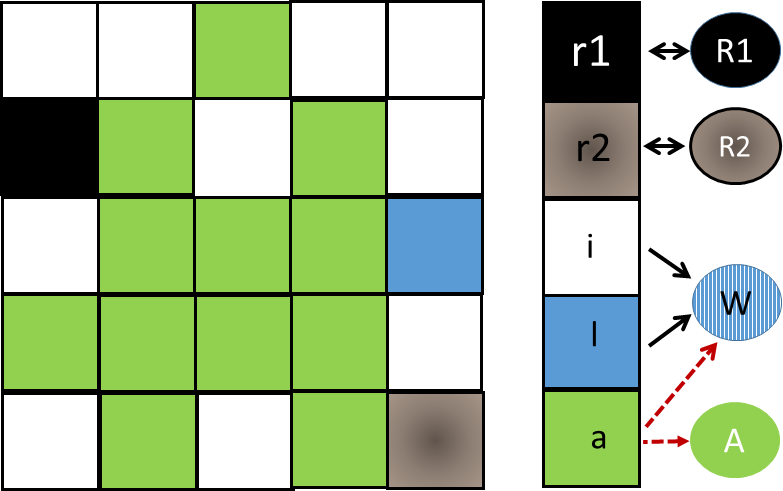}
\caption{Schematic representation of an REV for micro-scale of soil (left) and of dirty snow (right) showing the phases $r1,r2,i,l,a$ and how components $R1,R2,W,A$ are partitioned between these phases. In this paper the air phase is denoted by $g$, and we do not distinguish between $r1,r2$. See Examples~\ref{ex:soilm} and \ref{ex:snow}.}
\label{fig:porous}
\end{figure}

Porous media such as soils are modeled at the micro-scale called pore-scale or at the macro-scale called Darcy scale. At the Darcy scale with length scale $\abs{\Omega}=O( \mpunit{m})$ we work with $x\in \Omega$, where $\Omega$ is an open bounded porous reservoir. At the pore-scale  we work in an open bounded simply connected region $\omega(x)$ with characteristic length scale $\abs{\omega}=O(\mpunit{\mu m})$ so that $\epsilon \approx 10^6$. 

The soil domain $\Omega^{(soil)}$ is a composite material, made of components $R,W,A$ arranged in phases $r,l,i,g$, with the phases associated with the subdomains $\Omegar,\Omegal,\Omegai,\Omegag$ so that $\Omegas = \Omegar \ccup \Omegal \ccup \Omegai \ccup \Omegag$. 
Each component can be only in certain phases as shown in Figure~\ref{fig:porous}. Rock phase can be made of more than one rock component such as silica, quartz, calcite, clay, or organic matter. In this paper we focus on two subtypes of material, mineral and organic, which for simplicity we lump into one phase $r$ by averaging their properties.  

Here we assume that the porous medium is rigid, thus the interface between $\Omegar$ and $\Omegal \ccup \Omegag$ is fixed.  We also assume that the air pockets are immobile,  thus the interface between $\Omegal$ and $\Omegag$ is also fixed. When modeling fluids at the Darcy scale  it is not practical to try to account for these individual interfaces. Instead, one works with volume fractions defined at the micro-scale. 

The micro-scale is the pore-scale where at $y \in \omega$ we recognize the individual rock grains within $\omegar$, and the void space filled with fluids such as water and air $W,A$, and hydrocarbons, appearing in ice $i$, liquid $l$, or gas phases $g$ in $\omegap,p=i,l,g$. To make a connection between micro- and macro-scales in soils and fluid phases, we set $\omegar(x)=\Omegar \cap \omega(x)$ with its volume fraction $\eta_r(x) = \frac{\abs{\omega_r(x)}}{\abs{\omega(x)}}$ and similar notation for other phases. We also assume that in each REV $\omega(x)$ we have nearly identical (e.g., periodic) arrangements of rock phase, and we  define porosity $\eta(x)=1-\eta_r(x)$. 
However, we allow that the fluid phases in  
$\omega = \omegar \ccup \omegal \ccup \omegai \ccup \omegag$ change their arrangements, which is quantified by the phase volume fractions $\eta_p(x)$ depending on $x$. The phase fractions for mobile phases \cite{lake,BearCheng10} are defined as  volume fractions with respect to the volume of void space
$S_p=\frac{ \abs{\omega_p}} {\sum_{p={i,l,g}} \abs  {\omega_p} }$ and are also called saturations \cite{lake}. See Table~\ref{tab:porous} for $\Omegas$-specific notation.

The insight into the distribution of $\omegap, p=r,i,l,g$ is provided by imaging, e.g., micro CT \cite{PTISW}, or through modeling \cite{PTISW,PUS,PVB,APS,PHV}. 

%%%%%%%%%
\begin{example}[Multiscale nature of soil]
\label{ex:soilm}
We illustrate a soil REV with  a cartoon in Figure~\ref{fig:porous}. We lump the two rock types together so that $\omegar = \omega_{r1} \ccup \omega_{r2}$, and distinguish air as the only gas phase $p=a=g$. For soil REV $\omega \subset \Omegas$, we have $\eta= 14/25, S_l=11/14, S_i =2/14, S_a=1/14. $ For snow REV $\omega \subset \Omegasn$ we have $\eta=23/25, S_l=1/23, S_a=12/23, S_i = 10/23$. 
\end{example}

\begin{table}[htbp]
\footnotesize
\centering
\begin{tabular}{|c|l|}
\hline
Notation & {Description}
\\
\hline
$p \in \{i,l,g\}$ & {Subscript: phase $p$: solid ice $i$, or liquid $l$, or gaseous (mostly air)}
\\
\hline
$C \in \{W,R,A\}$ & {Superscript for component $C$: water $W$, rock grains $R$, air $A$}
\\
\hline
$(m) $&soil types characterized by pore sizes; $m \in \{nano,micro,meso, macro\}$
\\
\hline
$\omega,\omegam$& A pore-scale subdomain REV and of type $m$ within the REV
\\
\hline
$\omega_r = \bigcup_m \omega_r^{(m)}$, $\omegamr$& domain occupied by rock grains (in sub-material $\omegam$)
\\
$\eta_r=\sum_m \etam \etamr$
& volume fraction occupied by rock in REV 
\\
\hline
$\omega \setminus \omegar$& REV domain occupied by fluids in mobile and immobile phases
\\
$\eta=1-\eta_r=1 -\sum \etam \etamr$& porosity of soil, e.g., volume fraction of void space;
\\
$\satmp= \frac{\abs{\omegamp}}{1-\abs{\omegamr}}$ & {Volume fraction of phase $p$ in the void space of $\omegam$; $\forall m, \sum_{p} \satmp =1$. }
\\
$S_p= \sum_m \etam \satmp$&  {Volume fraction of phase $p$ in the void space of REV; $ \sum_{p=i,l,g} S_p =1$. }
\\
\hline
\end{tabular}
\caption{Notation for phases and components in $\Omegas$ and $\Omegasn$.}
\label{tab:porous}
\end{table}

%%%%
\subsubsection{Pore sizes}
\begin{figure}[htp]
\centering
\includegraphics[height=4cm]{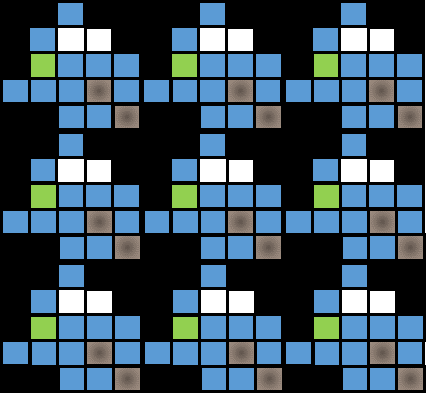}
\includegraphics[height=4cm]{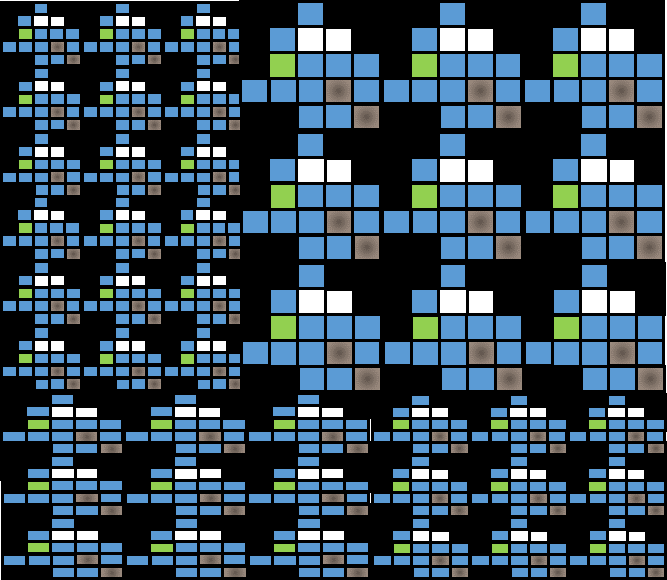}
\includegraphics[height=4cm]{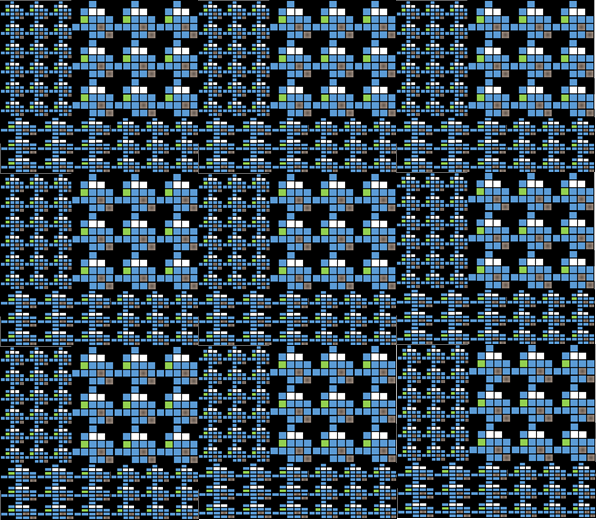}
\caption{Schematic picture of porous soil REV $\omega$ (phases are as in Figure~\ref{fig:porous}). From left to right: different pore sizes of domains $\omega \setminus \omegar$ (non-black voxels). Left: domain $\omega$ features only one average pore size about $1/4 diam(\omega)$; here $\epsilon=1/3$. Middle: $\omega$ features three distinct relative pore sizes $(r^{(m)})_m=(1/5, 1/6, 1/10)$ with volume fractions (0.5,0.2,0.3), respectively. 
Right: the soil agglomerate $\Omega=\bigcup_x \omega(x)$ is composed of heterogeneous subdomains, each a scaled copy of the domain in the middle, with $\epsilon=1/3$.}
\label{fig:porousm}
\end{figure}

When modeling heat conduction in soils in temperatures undergoing freezing and thawing, there is an additional complication. It is well known \cite{bookFGE} that the fluids in confined domains such as in the soil pores have different physical properties depending on the pore size; these range from less than $0.002~\mpunit{mm}$ in clay ($\eta \approx 0.5$)  to $0.05~\mpunit{mm}$ in silt ($\eta \approx 0.45$), and to $0.05$-$2~\mpunit{mm}$ in sand ($\eta \approx 0.32$); see \mpcitee{BearCheng10}{Pg. 74}. In particular, capillary pressures feature different curves for ``fine'' and ''coarse'' porous media \cite{BearCheng10}.
These pore distributions depend on several factors, including the type, shape and the variety of particles, and the degree of compaction of the soil. In top layers of soil, the pores can vary significantly, but the lower layers will have more small pores.  Interestingly, the imaging such as xray $\mu$-CT only ``sees'' a limited distribution of pore sizes due to the lack of resolution; see, e.g. \cite{Rooney22,Nimmo}. We illustrate the soil microstructure with multiple pore sizes with a cartoon in Figure~\ref{fig:porousm}. 

These same factors determine the freezing temperatures in soils with the tiniest crevices in nano-pores freezing the last; the smallest pores to contain some water molecules must be at least $1~\mpunit{nm}$ in size, thus the $\tfr^{(m)}$  can vary significantly between the $m$ types describing nano-, micro-, meso- and macro-pores. In particular,  the freezing in macro-pores such as in peat is similar to that in bulk water at a temperature close to $\tfr \approx 0\degc$, but the water films in nano-pores and small crevices may not even freeze at all in the range \eqref{eq:thetarange}, and  a typical freezing point depression is $\tfr \mpunit{^\circ C}$ $\in [-5, 0]$ \mpcitee{bookFGE}{Pg. 24}. This is explained, e.g., by the Gibbs-Thomson relationship or by phenomena of thermal regelation; we refer to  \cite{Rempel04,KW13,PVB} for a thorough discussion. 

For the purposes of this paper we distinguish the different soil sub-materials, and assume that every  REV $\omega(x)$ contains a considerable range of characteristic pore sizes as in Table~\ref{tab:soilmaterials}. 
In the end, the average liquid fraction $\chi$ in soils appears to have the support over the large range of temperatures in \eqref{eq:thetarange} unlike $H(\cdot)$ on $\R_+$ only for $\omegaw$.

%%%%%%%%%
\begin{table}
\footnotesize
\begin{tabular}{|c||c|c|c|}
\hline
&typical freezing temperature &volume fraction  &Effective model
\\
pore sizes in $\omegam$& $\tfr^{(m)}$ in $\omega^{(m)}$& &$\chis$, $\chires$
\\
\hline
\hline
nano-& $\tfr^{(m)}  \ll -5 $, & 
10 \% &$\chires>0$ 
\\
micro-& $\tfr^{(m)}  \approx -5$, & 
10 \% &
\\
meso-& $ \tfrm \approx -1$, & 
20 \% &
\\
macro- & $\tfr^{(m)} \approx 0$ & 
40 \% & $\chis<1$
\\
\hline
\end{tabular}
\caption{Example of properties of liquid to ice phase change properties for soil sub-domains $\omega^{(m)}$ depending on pore sizes,  and consequences for $\chires=\lim_{\theta \to -\infty} \Upsilon(\theta)$, and 
$\chis=\lim_{\theta \to 0} \Upsilon(\theta)$. See also Figure~\ref{fig:porousm} for illustration.
}
\label{tab:soilmaterials}
\end{table}
%%%

%%%%%%%%%%%
\subsection{Snow domain $\Omegasn$: description} 
\label{sec:materialsn}
Snow material is made primarily of $W$ and $A$ components which appear in $i,l$ and $g$ phases, respectively; see Figure~\ref{fig:porous}. Snow types depend on the atmospheric conditions and on the geographic region; in particular, the snow in the Arctic and Antarctica is different than that in high alpine, rural, or urban regions of lower latitudes; see \cite{colbeck,Jafarov2014,Sturm2002}. In particular, near urban areas or on glaciers one can also have dust particles embedded in the snow, and in alpine regions one can have cryophilic species such as Chlamydomonas nivalis algae which give the snow its pink color. Snow undergoes physical changes called metamorphism well described in geophysical literature \cite{Sturm2002}.  

At the micro-scale REV $\omega$ of snow domain $\Omegasn$ one recognizes the ice crystals in $\omegai$ surrounded by the air in $\omegag$, each of diameter $2$-$10~\mpunit{mm}$ diameter so that the REV size is $\abs{\omega(x)}\simeq 10^{-2}~\mpunit{m}$. Modeling of snow at the micro-scale is very complex and  requires advanced techniques to track the fluid phases \cite{Kaempfer}; the challenges are exacerbated by difficulty of imaging at subfreezing temperatures and the resulting paucity of data to develop and validate the models.  It is reasonable to model a film of liquid water around crystals, and/or droplets of undercooled water in deep snow, with $\chi \approx \hH(\theta-\tfr)$ similar to that in bulk water $\omegaw$.   
 
At macro-scale, $\Omegasn$ is a snow layer of typical thickness  $0.1$-$1~\mpunit{m}$. At this scale it makes sense to adopt the porous media nomenclature: assume $\eta \approx 1$, and use  phase volume fractions $S_p$ for fluid phases with $S_i+S_l+S_g=1$. 
Since the $g$ phase is made of $A$ and $W$ components, one requires the mass fraction $\mymu_g^W$ and should model its evolution. Finally, one can consider some sublimation phenomena when rapid increase of heat amount caused a change $i\to g$, but such a discussion is out of our scope.

%%%
\begin{example}[Snow, thermal properties at macro-scale]
\label{ex:snow}
For snow as a composite material at macro-scale,  we have $S_i+S_l+S_g=1$. 
Typically $S_g (x)\approx 0.9$ in the top fresh snow layers $x \in \Omega^{(snow,fresh)}$ where also $S_l \approx 0$ and there is no phase change. However, $S_g$ decreases with the snow depth and the level of metamorphism. For example, in deep, old, or corn snow   $x \in \Omega^{(snow,old)}$ we have $S_g(x) \in [0.2,0.5], S_l(x) \approx 0.1$ with $S_i \approx 0.4$. 
\end{example}

To complete a thermal model at macro-scale, we require a model for phase change of $W$ near $\tfr=0 \degc$  from $i$ or to $l$. Recognizing the difficulties with equilibrium brought up in Remark~\ref{rem:EQscales}, we postulate similar behavior as in $\Omegaw$.  

%%%
\subsubsection{Literature}
\label{sec:snowlit}
Snow cover plays an important role in the Arctic as an insulating interface between the atmosphere and the ground. For this reason heat transfer through snow should be included  as part of snow-soil model. In particular, the classical approach in 
\cite{NOAA} provides an energy and mass balance model for $\Omegasn$ involving the air-snow and snow-soil interfaces. The most important factor is the thermal conductivity $k(x)$ which is assumed to depend on snow density $\rho^{snow}(x)$ and other factors such as its type, the degree of compaction, and liquid content. In \cite{Jafarov2014} authors use an inverse modeling approach and estimate $k(x) \approx const$ treating $\Omegasn$ as part of coupled $\Omegas \ccup \Omegasn$ model which requires detailed data. In contrast, in \cite{LingZhang} a lumped model for surface energy treats $\Omegasn$ as a one dimensional interface, and adapts the boundary conditions for $\Omegas$ depending on various energy fluxes and solar radiation, air temperature, surface albedo and other factors. In \cite{PM24} we found that this particular model is highly sensitive to the many geophysical factors and their correlations. 

%%%%%%%%%%%%%%%
\subsection{Cryoconite domain $\Omegac$: description} 
\label{sec:materialsc}
Cryoconite is present in the Arctic, Antarctica, and high alpine regions \cite{Takeuchi2023}; $\Omegac$ is the ice or snow domain 
with grains of dust or dirt, e.g., charcoal or volcanic tuff, or organic matter blended with the grains; these are sometimes called cryoconite. In other words, $\Omegac$ is made of $R,A$ and $W$ in $r,g$ and $l,i,g$ phases, respectively. Due to increased solar absorption, the ice below the cryoconite melts faster than the surrounding ice, and cylindrical ``cryoconite holes'' with liquid phase of diameter  $5$-$145~\mpunit{cm}$ and depth $4$-$56~\mpunit{cm}$ appear \cite{Fountain,rozwalak2022,porazinska2004} covering up to $\approx 5$\% of surface.

To study cryoconite at the microscale, one recognizes that the mineral dust particles have typical diameter $60$-$250~\mpunit{\mu m}$   and the organic particles feature $130$-$400~ \mpunit {\mu m}$ \mpcitee{Takeuchi2018}{Tab.2}. Modeling at the micro-scale faces similar challenges to that in snow and soil: the REV size should be about $\abs{\omega(x)} > 10~\mpunit{mm}$ and it would include a variety of pore sizes and especially many small pores, as well as the macro-pores in which majority of the ice and snow resides.  

Regarding macro-scale of $\Omegac$, from the figures in \cite{Takeuchi2018} we estimate that the porosity $\eta \approx 0.6$ with $\eta_{ro} \approx 0.5$ and $\eta_{rm} \approx 0.1$. We also estimate $S_g \approx 0.4$ as in dense snow. In our model we allow $\chi$ to be similar to that in $\Omegas$ but with small $\chis$.  
We also have the volumetric fractions 
$\eta(S_i + S_g + S_l)$, and in $S_g$ we may have water vapor or air component, but we assume $\mymu_g^W=0$. 

Since we have some rock material, we allow some undercooled water; we expect to need $\chis$. Some of that water might be bound in organic matter (in bacteria). 

\begin{example}[Cryoconite, thermal properties]
\label{ex:cryo}
$\Omegac$ includes snow (ice and air) with grains of rock of mineral and organic type. 
The freezing temperatures $\tfr$ are $0\degc$ in the majority of ice/snow region. We also hypothesize that there may be undercooled water  $\tfr \approx -1$ in the crevices of some irregular rock grains.   
\end{example}

%%%%%%%%%%%%%%%
\subsubsection{Literature} 
\label{sec:cryo}
%%%
In \cite{rozwalak2022,{Takeuchi2018}}, the authors describe cryoconite as a mixture of mineral and organic material covering glacial ice, playing important roles in biogeochemical cycles and lowering the albedo of a glacier surface, since it is absorbing more solar radiation than the surrounding bare ice surface.  Cryoconite was found in 1870 \cite{WhartonMcKay1985} and is present in the Antarctic, the Arctic, and on glaciers in temperate regions adjacent to the Arctic. Recent classification in \cite{rozwalak2022} of cryoconite material classifies it as granules with $diam~\approx 0.3$-$1~\mpunit{cm}$ with regular or irregular edges, with organic matter 
proportion from $2$-$22$\% \cite{Takeuchi2002},  or as loose mineral without bio-aggregations. The properties of cryoconite such as reflectivity seem to depend strongly on the presence of organic matter and are linked to the presence of red algae. 

%%%%%%%%%%%%%%
\begin{table}[htp]
\footnotesize
\begin{tabular}{||l||l||l|l|l||l||}
\hline
Material $\Omega^{(m)}$&Porosity & Liquid $S_l$ & Ice $S_i$ & Gas $S_g$ & Vapor $\mymu_g^W$
\\
\hline
soil, air-free&$0< \eta <1$ & $0\ll S_l$ 
& $0 \leq S_i $ &0 &0 
\\
soil with large pores, air&$0< \eta \leq 1$ & $0\ll S_l$ 
& $0 \leq S_i $ &$0 \leq S_g$ &0 
\\
\hline
\hline
snow, dry/new&$\eta=1$ & $0\approx S_l $ 
& $0 \ll S_i $ &$0 \ll S_g $ &$0\leq \mymu_g^W$
\\
snow, wet/corn&$\eta=1$ & $0<S_l $ 
& $0 \ll S_i $ &$0 < S_g $ &$0\leq \mymu_g^W$ 
\\
snow, ``dirty''&$\eta \leq 1$ & $0<S_l $ 
& $0 \ll S_i $ &$0 < S_g $ &$0\leq \mymu_g^W $ 
\\
\hline
\hline
cryoconite&$0 \leq \eta\leq 1$ & $0 \leq S_l $ 
& $0 \ll S_i $ &$0 \leq S_g $ &$0\leq \mymu_g^W$ 
\\
\hline
\end{tabular}
\caption{Materials in $\Omega^{M}$ of interest to this paper. We have that $0 \leq S_p \leq 1$ and $\sum_p S_p=1$, but we annotate in a special way the importance of a particular phase. If a phase $p$ might or might not be present, we write $0 \leq S_p$, and if it is likely to be absent, we write $S_p \approx 0$. If phase $p$ is expected to be always present, we write $0 \ll S_p$. The mass fractions in column 6 are subject to the relative humidity constraint: $\mymu_g^W \leq \mymu_g^{W,*}(\theta)$. }
\label{tab:materials}
\end{table}

%%%%%%%%%%%%%%
\section{Models for heat equation with phase change} 
%%%%%
\label{sec:models}

Thermal energy conservation models in materials of interest in this paper require that we model the phase change which provides the key complication and challenge.  We recall the general background in Section~\ref{sec:thermo}. 
Thermal models with phase change are well understood for a single material, and we recall these for completeness in Section~\ref{sec:single}. Their mathematical structure has been less explored for complex composite materials which we address in Section~\ref{sec:composite}. In Section~\ref{sec:multi} we complete our outline of macro-scale model with phase change for a mixture of components. This will provide the framework for a unified generalized model to apply to $\Omegas, \Omegasn, \Omegac$ to be specialized in the subsequent Section~\ref{sec:newmodels}. 

%%%%%%%%%%%
\subsection{Thermodynamics background} 
\label{sec:thermo}
Consider a single material undergoing phase change between phases $p_1$ and $p_2$ at some temperature $\theta=\tfr$ at which such a change is possible. The phase behavior is denoted by phase fraction $\chi=0$ in phase $p_1$ where $\theta < \tfr$, and $1$ in phase $p_2$ where $\theta>\tfr$. The internal energy $w$ of this material 
 follows the differential law \mpcitee{BPV}{eq.~(2.5)}
\ba
\label{eq:dw}
dw = c d\theta + L d\chi
\ea
with temperature denoted by $\theta$ and the phase fraction denoted by $\chi$, where $\chi$ changes from 0 (phase $p_1$) to 1 (in phase $p_2$) during the phase change.  This law describes the change in internal energy due to the increments of the temperature while the fraction $\chi$ is fixed, and the latent heat amount required to change the phase fraction at some fixed phase change temperature. The specific heat $c$ may depend on $\theta$ or may be constant. The latent heat amount may depend on $\theta$ or be constant. The theory of thermodynamics allows to calculate $c_p,L_{p_1,p_2}$ depending on temperature, pressure, and composition of the substance; we refer to the extensive material and examples e.g. Chapter 4 in \cite{SvNA} and Chapter 9 on first-order phase transitions in \cite{Callen}.

For a particular given material or composite, one needs  to complete these steps:
\begin{algorithm} 
\label{alg:multiphase}
To identify thermal  properties for heat equation \eqref{eq:heat} with phase change
%%%
\begin{enumerate}
\item define $\chi$ and data for $L$,
\item provide values of $c$, and calculate $\cC$,
\item calculate $(\theta,\chi) \to w$ by integrating \eqref{eq:dw}, 
\item provide data on the conductivity $k$. 
\end{enumerate}
\end{algorithm} 
%%

%%%%%%%%%%%%%%%%%%
\begin{table}
\small
\begin{tabular}{|l||c|cc|c|}
\hline
Phase & Density $\rho_p$&specific heat $\cbar_p$ &$c_p$&conductivity $k_p$
\\
\hline
& $\mpunit{kg m^{-3}}$& $\mpunit{J/(kg \cdot K)}$ &$ \mpunit{J/ (m^3 \cdot K)}$&$\mpunit{W/(m \cdot K)}$
\\
\hline
$p=i$&$\rho_i$=$931$&$\cbar_i=2093 $
& $c_i=1.95 \times 10^{6}$ 
&$k_i=2.30$
\\
$p=l$&$\rho_l$=$1000$&$ \cbar_l=4184$
& $c_l = 4.19 \times 10^{6}$
&$k_l=0.58$
\\
$p=g$ (air) &$\rho_g^A$=$1.293$&  $\cbar_g^A \in [717,1000.5]$
&$c^A_g=0.927 \times 10^3$
&$k_g^A=0.026$
\\
$p=g$ (water vapor)
&$\rho^W_g$=$0.0048$&  $\cbar^W_g =1.9$
&$c^W_g=0.005 \times 10^3$
&$k_g^W=0.018$
\\
&$\mymu_g^W \in [0.001, 0.02]$  &&&
\\
\hline
\\
\hline
$p=r$ (sand, dry) &$\rho_r$=1420&$ \cbar_r=800$
&$c_r=1.13 \times 10^6$
&$k_r\in [0.27,0.34]$
\\
$p=r$ (clay, dry) &$\rho_r$=$1480$&$ \cbar_r \in [753,878]$
&$c=1.21\times 10^6$
&$k_r=0.25$
\\
$p=r$ (peat) &&
&$c_r=1.57 \times 10^6$
&$k_r=0.25$
\\
$p=r$ (silt)&
&
& $c_r=1.89 \times 10^6$& $k_r=2.92$
\\
\hline
\end{tabular}
\begin{tabular}{|l||l|l|l|}
\hline
$p_1 \to p_2$ & Temperature&Latent heat $\Lbar_{p_1,p_2}~\mpunit{J/kg}$ & $L_{p_1,p_2}~\mpunit{J/m^3}$
\\
\hline
Ice $i \to l$ liquid & $0\mpunit{\degc}$ & $\Lbar_ {i,l}^{(W)}  = 3.34 \times 10^5 \mpunit{J/kg }$ & $L_{i,l}^{(W)}= 3.06278 \times 10^8 ~\mpunit{J/m^3}$ \cite{RBC79}
\\
\hline
Liquid $l \to g$ gas & $\mpunit{100\degc}$& $\Lbar_ {lg}^{(W)}  = 22.65 \times 10^5 \mpunit{J/kg} $ & $L_{lg}^{(W)}= 22.65 \times 10^8 ~\mpunit{J/m^3}$ \cite{RBC79}
\\
\hline

\end{tabular}
\caption{Phase data at $\theta=0\degc$ from \cite{CRC, RBC79,bookFGE,LingZhang} and \mpcitee{WilliamsSmith89}{Pg. 90}. When no data or range is given, the $c_p$ value is the data directly from source, or the mean. }
\label{tab:phases}
\end{table}

%%%
\subsection{Model for phase change in a single material (pure substance)} 
%%%%%
\label{sec:single} 
We recall the water component $W$ in $\Omegaw$ with notes provided in Section~\ref{sec:omegaw}. We now follow 
\ralg{alg:multiphase} and complete the model \eqref{eq:heat} for the ice to liquid phase change with latent heat $L_{il}$ and from liquid to gas with $L_{lg}$. 
%%

%%%
\subsubsection{Ice-liquid} 
\label{sec:omegawil}
We assume the phase transformation  is at $\tfr=0\degc$. For the steps in \ralg{alg:multiphase} we need the 
liquid fraction $\chi = \onechar_{\Omegal(x,t)}$. Recalling $\gG$ defined in \eqref{eq:chizero} we have
\bsa
\label{eq:omegawil}
\ba
\label{eq:chidefST}
(\theta,\chi) &\in& \gG[\tfr,0,0].
\ea
%%%
It is also customary to write $\chi(\theta) \in  \hH(\theta-\tfr)$ or 
\ba
\chi =\left\{ \begin{array}{ll}
0, &\theta<\tfr,
\\
{[0,1]}, &\theta=\tfr,
\\
1, &\theta>\tfr.
\end{array}
\right.
\ea
%%%
We also have
\ba
c(x,t)&=& c_l \onechar_{\Omegal(x,t)}+ c_i \onechar_{\Omegai(x,t)} 
\\
k(x,t) &=& k_l \onechar_{\Omegal(x,t)}+ k_i \onechar_{\Omegai(x,t)}
\ea
The data on the phase properties is in Table~\ref{tab:Omegaw}. 
%%%%%%%%%
\begin{table}[htp]
\begin{tabular}{c||l|l||l|l||l|l||l|}
water $W$&
$\tfr$&
$L=L_{il}$&
$c_i$&$c_l$&
$k_i$&$k_l$&$\chi_l$
\\
\hline
&0 & 
$3.06 \times 10^8$
&
$1.90 \times 10^6$& $4.19 \times 10^6$&2.30&0.58&$\hH(\theta-\tfr)$
\\
\end{tabular}
\caption{
Phase data in  $\Omegaw$ for ice to liquid phase change: 
thermal properties \cite{RBC79,ZhangMichalowski15,bookFGE} converted to the volumetric quantities. 
}
\label{tab:Omegaw}
\end{table}

Integration of \eqref{eq:dw} gives
\ba
\label{eq:ccdefil}
\cC(\theta)&=&\left\{\begin{array}{ll}
c_i (\theta-\tfr),&\theta<\tfr,
\\
0, & \theta=\tfr,
\\
c_l (\theta-\tfr),&\theta>\tfr,\\\end{array}\right. 
\\
\label{eq:wdefomegaw}
w= \cC(\theta) +  L \chi
&\in&
\left\{ \begin{array}{ll}
c_i (\theta-\tfr), &\theta<\tfr,
\\
{[0,L]}, &\theta=\tfr,
\\
c_l (\theta-\tfr) +L, &\theta>\tfr.
\end{array}
\right.
\ea
Now we calculate 
\ba
w_* = w(\tfr^-) = \lim_{\theta \to \tfr^-}=0,\;\;
w^* = w(\tfr^+) = \lim_{\theta \to \tfr^+}=L.
\ea

\begin{remark}
\label{rem:omegawil}
The limits of the piecewise linear function $\cC:\R\to \R$ at $\pm \infty$ are $\pm \infty$. Thus, the relationship $(\theta,\chi) \to w$ is set valued on $\R^2$ but is a maximal monotone bijection from $\gG[\tfr,0,0]$ to $\R$. 
As a consequence, given any $w \in \R$, one can uniquely determine the pair $\psi=(\theta,\chi)$ and  invert the relationship \eqref{eq:wdefomegaw} $\psi \to w$ with values in $\gG[\tfr,0,0]$ as follows 
\ba
\label{eq:winvil}
\psi(w)  = 
\left\{ \begin{array}{lll}
\theta = \tfr + w/c_i, & \chi=0, &w < w_*=0,
\\
\theta=\tfr, &\chi=w/L, &w \in [w_*,w^*]=[0,L],
\\
\theta=\tfr + (w-L)/c_l, &\chi=1, & w>w^*=L.
\end{array}
\right.
\ea
\end{remark} 
\esa

We now have complete data in \eqref{eq:omegawil} for the heat equation \eqref{eq:heat}. Since the model involves the free boundary $\Gamma_{il}$ defined implicitly, as well as a discontinuity of $w$ across $\Gamma_{il}$, the  model \eqref{eq:heat} can be only  understood in the sense of distributions or analyzed after the so-called Kirchhoff transformation \cite{Show-monotone,VisBook}. 

%%%%%%%%%%%%%
\subsubsection{Phase change from liquid to gas of a single component}
\label{sec:omegawlg}
We briefly summarize the phase change from liquid to vapor for $W$ component at the boiling temperature $\tboil$ which may depend on the pressure, e.g., the boiling point for $W$ at lower pressures (such as in high elevations) is smaller than at high pressures, and might be  relevant for some environments. 
We set $\chi_g=  \onechar_{\Omegag(x,t)}$ and $(\theta,\chi_g) \in \gG[\tboil,0,0]$ analogous to the ice-liquid phase change
\bsa
\label{eq:omegawlg}
\ba
\label{eq:chideflg}
(\theta,\chi_g)(x,t)  &\in&
\left\{ \begin{array}{lll}
\chi_g=0, &\theta<\tboil, & x\in \Omegal
\\
\chi_g \in {[0,1]}, &\theta=\tboil, & x \in \Gamma_{lg}
\\
\chi_g = 1, &\theta>\tboil & x \in \Omegag, 
\end{array}
\right.
\\
c(x,t)&=& c_l \onechar_{\Omegal(x,t)}+ c_g \onechar_{\Omegag(x,t)},
\\
k(x,t) &=& k_l \onechar_{\Omegal(x,t)}+ k_g \onechar_{\Omegag(x,t)}.
\ea
We also have expressions for $\cC(\theta)$ analogous to \eqref{eq:ccdefil} which we complete with
\ba
w = \cC(\theta) +  L_{lg} \chi. 
\ea
\esa
The data is in Table~\ref{tab:Omegawlg}. The model has qualitatively the same behavior as that discussed in Section~\ref{sec:omegawil}, and one can set-up an inverse relationship analogously to \eqref{eq:winvil}. 

%%%
\begin{table}[htp]
\begin{tabular}{c||l|l||l|l||l|l||l|}
water $W$&
$\tboil$&
$L_{lg}$&
$c_l$&$c_g$&
$k_l$&$k_g$&$\chi_g$
\\
\hline
&$100\degc$ & 
$ 22.65 \times 10^8$
& $4.19 \times 10^6$
&$1.3 \times 10^3$ &0.58&0.018&$\hH(\theta-\tboil)$
\\
\end{tabular}
%%%
\caption{Phase data in  $\Omegaw$ for liquid to vapor phase transformation. 
}
\label{tab:Omegawlg} 
\end{table}
%%%

%%%%%%%%%%%
\subsubsection{Phase change from liquid to gas for water component in the presence of other components}
In practical applications of interest in this paper the gas phase may include not just the water vapor but also other gases.  

In $\Omegag$ we have the air component $C=A$ and the water component $C=W$.  
Air does not change phase when \eqref{eq:thetarange} holds, but the water may change phase as described in Sections~\ref{sec:omegawil} and \ref{sec:omegawlg}. 
However, now $\Omega_g$ contains two components, and the model to describe the state of the air-water vapor mixture is  determined by the mass fraction $\mymu_g^W$, typically expressed with data on relative humidity. In this case the change of energy is described by 
$dw = c d\theta + L^W_{lg} d \chi_g \mymu_g^C$; see, e.g.  \mpcitee{Garg08}{eq. 30, 35} for a more complicated expression.
For gas, advection and buoyancy are important, but a discussion is out of our scope. 

In soils when \eqref{eq:thetarange} holds,  is unlikely that the phase change $l \to g$ in $\Omegas$ would occur, thus we set $\mymu_g^W=0$ in $\Omegas$. In $\Omegasn$ this phase change should be considered along with sublimation, which is out of present scope.

%%%%%%%%%%%%%
\subsection{Model for phase change in a composite materials and upscaling} 
\label{sec:composite}
We focus now on $\Omegas$ as a composite material which presents a significant challenge with respect to, e.g., to $\Omegaw$. This complexity is sufficient to connect to the generalized model to be developed for all materials including $\Omegasn$ and $\Omegac$. We follow \ralg{alg:multiphase} for the micro-scale which are then upscaled to the macro-scale and relate to \eqref{eq:dw}. 

As explained in Section~\ref{sec:materialss}, $\Omegas$ is made of multiple sub-materials all present within each REV $\omega(x)$, each with separate thermal properties indicated by subscripts $^{(m)}$ extending  Table~\ref{tab:phases}, with each portion $\omegam$ of the REV $\omega$ made of multiple phases $\omegap,p=r,l,i,g$. A modeler may be interested in tracking down these phase boundaries at the microscale \cite{BPV,Kaempfer,PVB,PHV}.  

For a model at micro-scale, we require $\chi^{(m)},c^{(m)},k^{(m)},L^{(m)}$ for each material, called {Model $\omega$-extended} in \cite{PVB}. In particular, $\chi^{(m)}=\hH(\theta-\tfrm)$ depend on $\tfrm$ which in soils depend on the pore-size. For the materials that do not undergo phase change in \eqref{eq:thetarange} we set $\tfrm=0$, and $L_{p_1,p_2}^{(m)}=0$. For $W$ with $p=i,l$ we assumed identical properties $c_p^{(m)},k_p^{(m)},L_{il}^{(m)}$ across all $m$, and we generalized \eqref{eq:omegawil} to
\ba
\label{eq:alpham}
y \in \omegam: w&\in& \begin{cases}
c_i(\theta - \tfr^{(m)}), &  \theta < \tfr^{(m)},
\\
[0, L_{il}^{(m)}], & \theta = \tfr^{(m)},
\\
c_l(\theta - \tfr^{(m)}) + L_{il}^{(m)}, & \theta > \tfr^{(m)}.
\end{cases}
\ea
For each $m$ and $\omegam$, we can also write $c$ and $k$ as well as the inversion formula for \eqref{eq:alpham} entirely analogously as in Section~\ref{sec:omegawil}.

To develop the macro-scale model we follow \cite{Vis2007,PVB}. By homogenization we construct Model-$\Omega$-extended, with the effective properties 
\bas
\chieff, \ceff, \cC_{eff}, \keff.
\eas
We start with  the first three which are piecewise constant and/or piecewise linear with changes at each $\tfrm$, and later address $\keff$.

\subsubsection{Effective $\chieff,\ceff$}
\label{sec:ceffm}
 In particular, we have
\bsa
\label{eq:eff}
\ba
\chi_{eff}(\theta) = \sum_{m=1}^{\nmat} \eta^{(m)} \hH(\theta-\tfr^{(m)})
\ea
which is piecewise constant and takes jumps at each $\tfr^{(m)}$ of magnitude $\Delta \chi_{eff}(\tfr^{(m)})=\eta^{(m)}$. 
In turn, the slopes of $\cC_{eff}(\theta)$ are given by 
\ba
\label{eq:ceff}
c_{eff}(\theta)\vert_{
(\tfr^{(m-1)}, \tfr^{(m)})
}
=
c_l \sum_{k=1}^{m-1} \eta^{(k)}
+
c_i\sum_{k=m}^{\nmat} \eta^{(k)}
= c_l \chi_{eff}(\theta) +c_i (1-\chi_{eff}(\theta)).
\ea
\esa
At each $\tfr^{(k)}$ the singular portion of $w(\theta)$ jumps by $L_{il}\Delta^{(k)} \chi_{eff}$. 
%%%

The formulas \eqref{eq:eff} are quite informative but are not easy to work with due to the multiple jump points. Instead, we consider an approximation by a smooth $\wt{\chieff}$ from which other formulas follow.
%%%
\subsubsection{Fit $\chieff$ to $\wt{\chieff}$ }
\begin{figure}[htp]
\centering
\includegraphics[height=4cm]{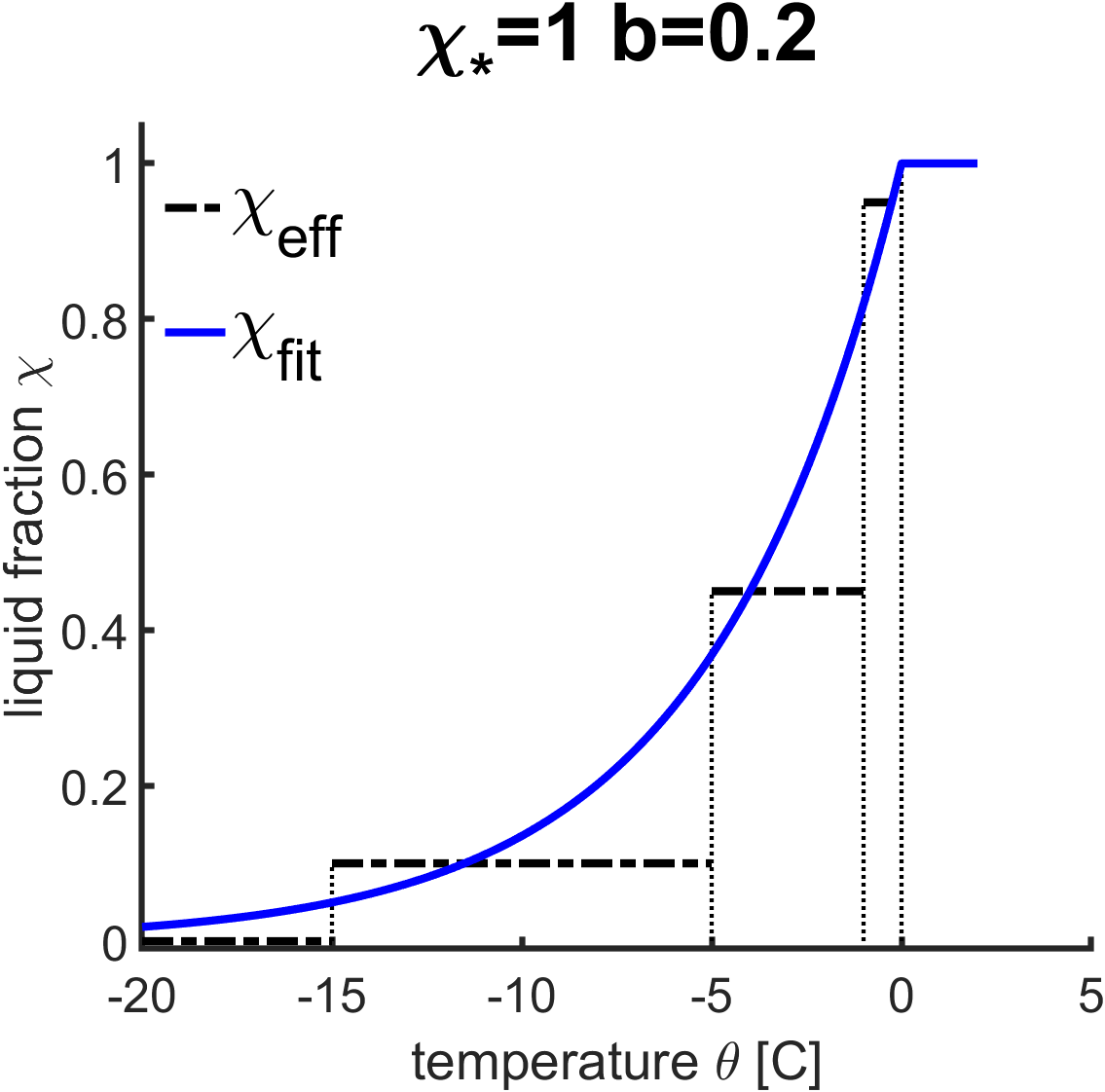}
\includegraphics[height=4cm]{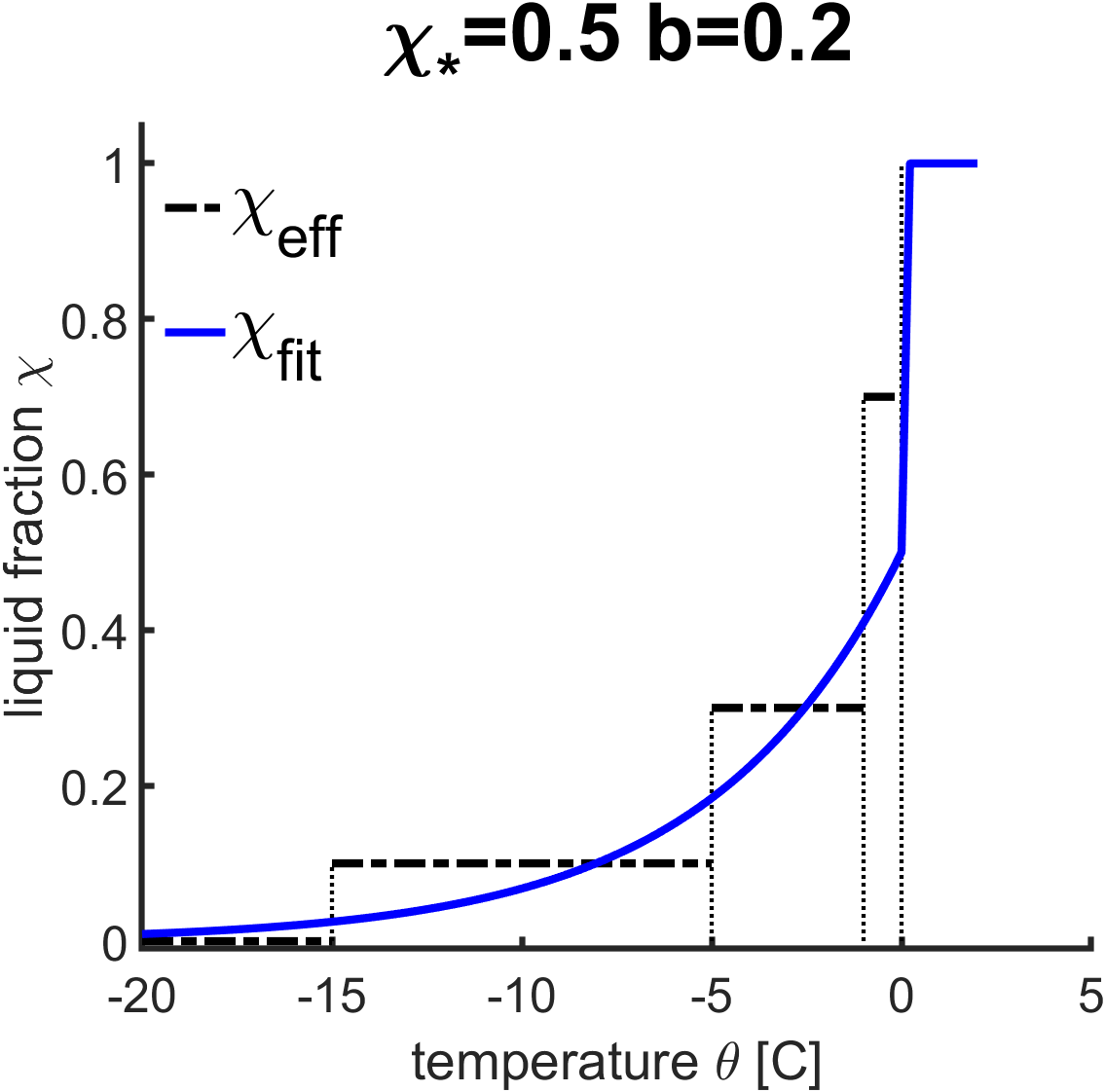}
\caption{Plots of $\chieff$ and $\wt{\chieff}$ where $\wt{\chieff}$ is continuous (left) and multivalued  (right)}
\label{fig:chi}
\end{figure}

The piecewise $\chieff$ can be fit to some smooth $\widetilde{\chieff}$.  In another view, $\widetilde{\chieff}$ is a continuum limit of the discrete $\chieff$ for $(\chi^{(m)})_m$ with $\nwmat \to \infty$. See Figure~\ref{fig:chi} for illustration. 

%%%%%%

We note that in \cite{PVB,PHV} we only considered nano- to meso-pores with $\tfrm\ll 0$ and with small $\etam, m=\nwmat \approx 0$. These resulted in a reasonable fit with $\wt{\chieff}$ continuous at $\theta=0$, i.e., with 
$
\wt{\chieff}(0^-)=1=\wt{\chieff}(0^+).
$
%%

%%%%
\subsubsection{Empirical $\chiemp$}
\label{sec:SFC}
Most interesting is that for soils, our fitted $\wt{\chieff}$ resemble the empirical $\chiemp$ called SFC (soil freezing curves) which satisfy 
\bas
\chiemp(0^-)=\chis < 1, \; \chiemp(0^+)=1,\;\;
\lim_{\theta \to -\infty}\Upsilon(\theta)=\chires>0.
\eas
These satisfy \rass{ass:upsilon} except for $\chires=0$ which we now ignore wlog. The SFC  feature a variety of formulas for $\Upsilon$ from power to exponential models for different soil types \cite{JAWheeler73,Osterkamp1987,Michalowski93,LingZhang,Kozlowski2007,RenVanapalli17,KW13}. Some have $\chis<1$ which correlates to the presence of macro-pores such as in soils dominated by organic matter, e.g., peat. For example, estimating from peat data  in \mpcitee{LingZhang}{Fig. 2} we find $\chis =0.16$ in peat, and $\chis = 0.40/0.55$ in silt.

\begin{remark}[Generalized $\chi$]
\label{rem:chi}
In the unified generalized models in this paper we postulate that the liquid fraction in the composite materials of interest at the macro-scale is defined by \rdef{def:chi} and satisfies  \rass{ass:upsilon} which we recall for simplicity 
\ba
\label{eq:chi} 
\chi(-\infty)=0; \chi(0^-)=\chis \leq 1, \; \chi(\R_+ \cup \{0^+\})=1.
\ea
In particular, \eqref{eq:chi} is satisfied by the empirical $\chiemp$ and by the upscaled fitted models $\wt{\chieff}(\theta)$ adequate for $\Omegasn$. We also note that the bulk water liquid fraction \eqref{eq:chidefST} satisfies \eqref{eq:chi} with $\chis=0$, which is useful for modeling $\Omegasn,\Omegac$. 
\end{remark}

%%%%%%%%%%%%%%%%%%%%%
%
In upscaled models with nonlinearity, we tacitly assume that the effective model, e.g., $\chi=\chi(\theta)$ is meaningful with its dependence on the average temperature at the macro-scale.  While this assumption resembles the empirical approaches, one should proceed with caution, as we addressed in Remark~\ref{rem:EQscales}. We plan to address the lack of equilibria and hysteresis following our work in \cite{PS24}, but a comprehensive discussion is out of scope here.

%%%
\subsubsection{Upscaling $k$ to $\keff$ when multiple phases are present} 
\label{sec:keff} 

First, we acknowledge that the literature postulates formulas for $\kemp$ applying \eqref{eq:aveag} $[A]$ or $[G]$ to phase fractions and incorporating $\eta$ and $\chiemp$. In other words, these approaches accept that $\keff$ depends only on the proportion $\chi$ of the liquid/ice, rather than on the arrangement of phases at the microscale. Regarding $\kemp(\theta)$, many authors use $\ave{k}^{[G]}$  \pcite{Osterkamp1987,JafarovMarchenkoRomanovsky12, LingZhang, Michalowski93, ZhangMichalowski15,bookFGE} but $\ave{k}^{[A]}$ is used in \pcite{LiuMaghoul19}.

Meanwhile, the rigorous upscaling \eqref{eq:keff} defined in Section~\ref{sec:upscale-onephase} to $\keff=\ave{k}^{[UP]}$ requires solving an auxiliary elliptic equation; in \cite{PVB} we show the differences between $\keff=\ave{k}^{[UP]}$ and $\keff=\ave{k}^{[A]}$ or $\keff=\ave{k}^{[G]}$. At the same time, calculation $\keff=\ave{k}^{[UP]}$ requires knowledge of geometry of phase distribution at the pore-scale such as in Figure~\ref{fig:porousm}. The issue is that such distributions require experiments or imaging, both of which are prohibitively complex. In addition, there may be many possible geometries for any fixed average liquid volume fraction $\chi$, and an entirely random geometry that fits a given $\chi$ might be also completely unrealistic. 

To avoid these conceptual and practical difficulties, in \cite{PUS,PHV} we proposed surrogate models for  $\widetilde{\keff}(\theta)$ based on constrained random realizations. We also showed that $\ave{k}^{[G]}(\theta)$ based on $\wt{\chieff}(\theta)$ is a good approximation to $\keff(\theta)$. We return to the calculation for $\keff$ for our generalized model in examples in Section~\ref{sec:NUSFF}.  

%%%
\subsection{Macro-scale thermal model with phase change for a mixture of components} 
\label{sec:multi}

Having discussed rigorous upscaling from the micro- to macro-scale, now we return to the background in Section~\ref{sec:thermo} and \ralg{alg:multiphase} with the aim to generalize the discussion in Section~\ref{sec:composite} 
with the particular materials $\Omegas,\Omegasn,\Omegac$ featuring different restrictions on these as shown in Table~\ref{tab:materials}, and with the material data in Table~\ref{tab:phases}. 

In particular, we shall assume gas phase with $S_g =const \ge 0$, but we do not account for the evolution and movement of the phases due to pressure difference. When integrating  the expression  \eqref{eq:dw} for multiple components and phases, we extend the expression $cd\theta$ to allow phase fraction dependent properties $c$. Separate consideration is given to the term $Ld\chi$ which describes how the phases are distributed between materials.  At the macro-scale, we work with phase fractions  rather than pointwise near phase interfaces, and must provide macro-scale phase specific internal energies in the expression \mpcitee{lake}{Section 2, 2.3-4} 
\ba
\label{eq:alpha-mixture}
w = (1-\eta)u_r + \eta S_i u_i + \eta S_l u_l + \eta S_g u_g.
\ea
This approach is followed in hydrological literature even without being explicitly or universally acknowledged.  It might be useful to see   how certain terms in \eqref{eq:alpha-mixture} are handled and dropped including the contributions associated with the pressure-volume work and the advection of latent heat; we refer to \cite{PHV}.  Below we continue with details on \eqref{eq:alpha-mixture}.

%%%
\subsubsection{Internal energies for phases} 
\label{sec:internal}
Internal energy definition depends thus on $S_p,\rho_p$, and is determined based on specific heat (heat capacity coefficients) $c_p$ and latent heat amounts $L_{p_1p_2}$   and depending on $\theta,\chi,\mymu_p^C$.  

\paragraph{\bf Solid phases} For phases $p=r,i$ which contain only a single component and do not undergo phase change, we have $u_p=\rho_p U_p$ and $dU_p = \cbar_p d\theta$, where $\cbar_p$ is the isobaric specific heat (assumed constant here). Thus $u_p=c_p(\theta-\tref{p})$, where $c_p = \rho_p \cbar_p$ as in Table~\ref{tab:phases}, and $\tref{p}$ is some reference point for the phase $p$; below we set $\tref{r}=0,\tref{i}=\tfr$. 

\paragraph{\bf Liquid phases} For the phase $p=l$ which contains only the water component, we account for the latent heat of the phase change $i \to l$, and $u_l = c_l (\theta-\tfr) +L_{il}$, with the fusion (melting/freezing) latent heat $L_{il}=\Lbar_{il} \rho_i$.  We assume that $S_l$ is given by $\chi(1-S_g)$ and $\chi(\theta)$ is one of the curves satisfying \eqref{eq:chi}. 

\paragraph{\bf Gas phase} 
For phase $g$, situation is more complicated, since it may contain component $A$ as well as component $W$ partitioned according to the mass fractions $\mymu_g^A,\mymu_g^W$, with $\mymu_g^A+ \mymu_g^W=1$. 
First, we see from Table~\ref{tab:phases} that one can approximate $c_g \approx c_g^A$, since $\mymu_g^W\leq 0.02$ is small, and even though the specific heats $\cbar_g^A \neq \cbar_g^W$, we have $\rho_g^A \gg \rho_g^W$. Thus
\ba
u_g \approx c_g (\theta-\tzero), \; \theta<\tboil, \;  c_g \approx c^A_g 
\ea
Second, the component $W$ might undergo vaporization/condensation phase change $l \to g$  at the temperature $\tboil$, but the air $A$ does not change phase when \eqref{eq:thetarange}.  We have 
\bas
u_g \approx c_g (\theta-\tzero) +L^W_{lg} \mymu_g^W, \theta \geq \tboil,  \; \; L^W_{lg}=\Lbar^W_{lg} \rho_l ,\;  c_g \approx c^A_g, \; \mymu_g^W \leq \mymu_g^{W,*}=0.005.
\eas
Here $\mymu_g^{W,*}$ is the maximum vapor fraction possible at the given temperature and pressure. We stress that even though $L^W_{lg} \gg L_{il}$, with $\mymu_g^W \approx 0.005$ we have $L_{lg} = \mymu_g^W L^W_{lg} %\approx 0.1 
\ll L_{il}$. In addition, due to the large density difference between $\rho_g^W \ll \rho_g^A$, the water vapor has large buoyancy, thus at phase transition to vapor it will likely be transported away. Finally,
near the triple point of water \cite{CRC} one can also consider a direct transformation $i \to g$ called deposition/sublimation, but one can model this change by a rapid change $i \to l \to g$. In the end, we assume $S_g \approx const$, discussion of the more general case is outside our scope. 

\begin{table}[htbp]
\footnotesize
\centering
\begin{tabular}{|c|c|l|}
\hline
&Notation & {Description}
\\
\hline
Any domain &$\mymu_i^{W}=1$ & Ice phase contains only $W$ component
\\
&$\chi$& Volume fraction of phase $l$ of component $W$ relative to $l$ and $i$: $\chi=\frac{ \abs{\omegal} }
{ \abs{\omegal \ccup \omegai}}$ 
\\
\hline
$\Omegas$&$\mymu_r^{(soil), R}=1$ & In soil, rock phase contains only $R$ component
\\
&$\mymu_l^{(soil), W}=1$ & In soil, liquid phase contains only $W$ {component}
 \\
&$\mymu_g^{(soil), A}=1$ & In soil, gas phase contains only air
\\
\hline
$\Omegasn$&$\mymu_g^{(snow), A} \leq 1$ & In snow, gas phase contains air and water vapor
\\
\hline
\end{tabular}
\caption{Components and phases in $\Omegaw,\Omegas,\Omegasn$. Properties in $\Omegac$ are similar to those in $\Omegasn$.}
\label{tab:parameters}
\end{table}
%%

%%%%%%%
\subsection{Summary: thermal model at macro-scale} 
Now we complete \ralg{alg:multiphase} integrating the expression \eqref{eq:dw} for the phase distribution in 
\eqref{eq:alpha-mixture}, with some known $\chi$, and with internal energies calculated as in Section~\ref{sec:internal}. 

We obtain
\ba
\label{eq:accumulation}
w = (1-\eta)  c_r \theta +\eta c_i S_i (\theta-\tfr)
+ \eta S_l (c_l (\theta-\tfr) +L_{il}) + 
\eta S_g (c_g (\theta -\tzero)).
\ea
To close, we must know how $S_i, S_l$ and $S_g$ depend on $\theta$, and we make some approximations and assumptions.   

The main assumption is to relate $S_l$ and $\chi$. Let $\chi=\frac{ \abs{\omegal} }
{ \abs{\omegal \ccup \omegai}}$ satisfy \eqref{eq:chi} to play a role of generalized liquid fraction from Remark~\ref{rem:chi}. Next we assume that this liquid fraction is in equilibrium with temperature, and it is straightforward to associate the liquid and ice fractions follow from  
\ba
\label{eq:chisl}
S_l = \chi (1-S_g); \;\;\; S_i = (1-\chi)(1-S_g).
\ea
%%%
After this major decision is made, we substitute to \eqref{eq:accumulation}, and complete the steps in   \ralg{alg:multiphase}.  
We show how this works for the 
simplified case known 
from \mpcitee{BPV}{eq. 5.4} and \cite{PVB}. 

%%%%%%
\begin{example}[Soil with $S_g=0$ (no gas) and $\chis=0$ (no macro-pores)]
\label{ex:soilP} 
Let some $\chi$ satisfy \eqref{eq:chi}, $\tfr=0$, $\chis=1$, and $S_g=0$. Then $S_l=\chi$. 
With the notation introduced in \eqref{eq:ave} we have 
\bas
c = \left[ \begin{array}{cc}
1-\eta & c_r
\\
\eta \chi & c_l
\\
\eta (1-\chi) & c_i
\end{array}
\right], \; 
k = \left[ \begin{array}{cc}
1-\eta & k_r
\\
\eta \chi & k_l
\\
\eta (1-\chi) & k_i
\end{array}
\right]. 
\eas
We calculate the useful weighted arithmetic averages
\bsa
\label{eq:soil}
\ba
\cuf = \ave{ 
\left[ \begin{array}{cc}
1-\eta & c_r
\\
\eta & c_l
\end{array}
\right]}^{[A]},
\cfr = \ave{ 
\left[ \begin{array}{cc}
1-\eta & c_r
\\
\eta & c_i
\end{array}
\right]}^{[A]}.
\ea
Thus $c$ at the pore-scale depends on the position within the phase regions, but its macro-scale average $c_{ave}= \cuf \chi + \cfr(1-\chi)$ depends on the liquid volume fraction $\chi$ 
\ba
c_{ave}= 
\ave{ 
\left[ \begin{array}{cc}
1-\eta & c_r
\\
\eta \chi & c_l
\\
\eta (1-\chi) & c_i
\end{array}
\right]}^{[A]}
= \cfr + (\cuf-\cfr) \chi.
\ea
After some calculations from $w = (1-\eta)u_r + \eta S_i u_i + \eta S_l u_l$ we get 
\ba
\label{eq:alphasoil}
w= \int_0^{\theta} c_{ave}(v) dv + L\eta \chi(\theta)
= \cC(\theta) + L\eta \chi(\theta),
\ea
%%%
where  the regular part of the energy \ba
\cC(\theta) = \int_0^\theta c_{ave}(v)dv.
\ea
With these calculations, since $\chi$ is continuous, $\theta \to w$ is a bijective function. The inversion formula is not explicitly available algebraically, and we discuss it in Section~\ref{sec:newmodels}. 

In turn, the heat conductivity is obtained from $k$ by upscaling to get $\ave{k}^{[G]}$ or $\ave{k}^{[UP]}$. 

Taking $\partial_t w$ which is defined when  $\theta \neq \tfr$ we have
%%%
\ba
\partial_t w =  \partial_t( \cC(\theta) + \eta L_{il} \chi ) = c_{ave}(\theta)  \partial_t \theta
+ \eta L_{il} \partial_t \chi
 = c_{app}(\theta) \partial_t \theta, \;\;\theta \neq \tfr.
\ea
\esa
We see the term $c_{app}(\theta) =c_{ave}(\theta) +\eta L_{il} \frac{d \chi}{d\theta}$ popular in the permafrost soil literature. 
\end{example} 

This example is extended to $\Omegas,\Omegasn,\Omegac$ in Section~\ref{sec:newmodels}. 

%%%%%%%%%%%%%%%%%%%%%
\section{Thermal models for soils with macro-pores and snow}
\label{sec:newmodels}
We now generalize Example~\ref{ex:soilP}  to the soils $\Omegas$ with $\chis\neq 1$, $S_g \neq 0$, and to all materials $\Omegasn,\Omegac$. It is helpful to organize the steps systematically. 

%%%%
\begin{algorithm}
\label{cor:stepsP}
Collect properties for composite material and phase change $i\to l$ from Table~\ref{tab:parameters}.
%%%
\begin{enumerate}
\item
\label{step:data} 
Identify $\tfr,\chis,S_g,\eta,\mymu_g^W$, and $c,k$ in \eqref{eq:cksoilg}, e.g, from Table~\ref{tab:phases}.
\item 
\label{step:upsilon}
Select the constitutive property  $\chi$ in \eqref{eq:chis}  with $\Upsilon$ satisfying \rass{ass:upsilon}.
\item For a given $c,\eta, S_g$ calculate $\cuf,\cfr$, and set 
\bsa
\label{eq:cregular}
\ba
\label{eq:cave}
c_{ave} = \cfr + (\cuf-\cfr) \chi.
\ea
\item For $\chi$ selected in step~\ref{step:upsilon}, integrate formally $\chi$ to get
\ba
\label{eq:cc}
\cC(\theta)=\int_{\tfr}^{\theta} c_{ave}(v)dv=
\left\{
\begin{array}{ll}
 (\cuf-\cfr) \int_{\tfr}^{\theta} \Upsilon(v)dv  + \cfr(\theta-\tfr), &
\theta < \tfr
\\
\cuf(\theta-\tfr), & \theta\geq  \tfr.
\end{array}
\right. 
\ea
\esa
Note that $\cC$ is continuous and, in particular, at $\theta=\tfr$, $\cC(\tfr)=0$.
\item Collect $w = \cC(\theta) + L_{il}\eta \chi $.
\item 
\label{step:inversion} 
Set up the inversion formula for the relationship $w \leftrightarrow \psi=(\theta,\chi)$. 
\item \label{step:capp} If possible, calculate $c_{app}$ for use in sequential models with $\theta$ as a primary unknown.
\item If step \ref{step:capp} is not possible, set up $w$ as primary unknown. 
\item Set up the calculation $\ave{k}^{[G]}$ or $\ave{k}^{[UP]}$ from hybrid or reduced models. 
\end{enumerate}
\end{algorithm}
%%%%%%%%%%%%%%%
We make these steps explicit for each of the materials. 

\subsection{Generalized model for soil $\Omegas$ with $S_g=const, \chis \leq 1$} 
%%%%%%%%%%%%%%
\subsubsection{Case $S_g=const, \mymu_g^W=0$ in $\Omegas$ with $\chis=1$}
We choose 
\bsa
\label{eq:soilsg}
\ba
\label{eq:chidefP}
\chi =
\left\{ \begin{array}{ll}
1, &\theta>\tfr,
\\
\Upsilon(\theta-\tfr), &\theta<\tfr.
\end{array}
\right.
\ea
We also account for trapped air bubbles which do not move. 
We obtain $w$ 
%%%
%%
\ba
\label{eq:soilg}
w = \cC(\theta) + \eta (1-S_g)  L_{il} \chi,
\ea
with $\cC(\theta) $ similar to  \eqref{eq:cc} and $c_{ave}=\ave{c}^{[A]}$ but with new definitions
\ba
\label{eq:cufr}
\cuf = (1-\eta)c_r + \eta S_g c_g + \eta(1-S_g) c_l,
\cfr = (1-\eta)c_r + \eta S_g c_g + \eta(1-S_g) c_i.
\ea
%%%
The heat conductivity
$\ave{k}^{[UP]}$ or $\ave{k}^{[G]}$
is obtained by upscaling or geometric averaging of
\ba
\label{eq:cksoilg}
c=\left[ \begin{array}{lc}
1-\eta & c_r
\\
\eta (1-S_g) \chi & c_l
\\
\eta (1-S_g) (1-\chi) & c_i
\\
\eta S_g & c_g
\end{array}
\right], 
\; 
 k = \left[ \begin{array}{cc}
1-\eta & k_r
\\
\eta (1-S_g) \chi & k_l
\\
\eta (1-S_g) (1-\chi) & k_i
\\
\eta S_g & k_g
\end{array}
\right]
.
\ea
\esa
For these we can calculate $c_{app}(\theta) =c_{ave}(\theta) +\eta (1-S_g) L_{il} \frac{d \chi}{d\theta}$.  

Knowing $S_g$ the inversion formula $w \to \psi=(\theta,\chi)$ is possible with a local nonlinear solver; see an example in Section~\ref{sec:ccexample}. After $w \to \theta$ is known, we can calculate $\chi=\chi(\theta)$. 

%%%%%%%%%%%%%%
\subsubsection{Case $S_g=const, \mymu_g^W=0$ in $\Omegas$ with $\chis< 1$}
\label{sec:soils} 
This case blends the formulas for $\Omegaw$ with those from $\Omegas$. We assume that $0\leq \chi(0^-)=\chis <1$. The expression $\eta (1-S_g)  L_{il} \chi$ includes a regular portion and an irregular portion in $(\theta,\chi) \in \gG[\tfr,\chis,\Upsilon]$
\bsa
\label{eq:defPs}
\ba
\label{eq:chidefPs}
\chi (\theta) \in
\left\{ \begin{array}{ll}
\Upsilon(\theta-\tfr), &\theta<\tfr,
\\
{[}\chis, 1{]}, & \theta = \tfr,
\\
1, &\theta>\tfr.
\end{array}
\right.
\ea
Other than that, we follow calculations as in \eqref{eq:soilg}--\eqref{eq:cksoilg}.

\begin{remark}
Since $\chi$ is not continuous, $w$ is multivalued at $\theta=\tfr$, and we cannot calculate $c_{app}$ or $\partial_t w$.  
\end{remark}

For the inversion formula, it is useful to calculate $w_*,w^*$ 
\ba
w_* = w(\tfr^-)=\eta (1-S_g)L_{il}\chis, \; w^* = w(\tfr^+) = \eta(1-S_g)L_{il}.
\ea 
The inversion formula is only partially explicit
\ba
\label{eq:needsolver}
(-\infty,w_*) \ni w &\to& \psi= (\theta,\chi), \;\; {\text{requires\ a\ solver}}, 
\\
(w_*,w^*) \ni w &\to& \theta=\tfr, \chi = \frac{w}{\eta(1-S_g)L_{il}},
\\
(w^*,\infty) \ni w &\to& \theta=\tfr+\frac{(w-w^*)}{\cuf}, \chi=1
\ea
\esa
In \eqref{eq:needsolver} we mention local nonlinear solver. We develop an example in Section~\ref{sec:ccexample} and the solver is in Section~\ref{sec:nsolver}. 

\subsection{Generalized macro-scale model for snow and cryoconite}
We recall the characteristics of these materials from Table~\ref{tab:materials}, and the  properties in Table~\ref{tab:parameters}. 
Most significant is that we work with temperatures at which $S_i$ and $S_g$ are nontrivial. Also, the gas phase contains $A$ which does not change phase, and $W$ which may change phase from $i$ to $l$ to $g$. 
We focus on the phase change $i\to l$, and assume $S_g=S_g(x)$ is known, while the temperatures $\theta<\tboil$, thus $\mymu_g^W \approx const$.  In the future we plan to account  also for the change of $W$ phase from liquid or directly from ice to vapor, which would require handling
$
L_{lg} \mymu_g^W S_g
$
but a comprehensive discussion as well as accounting for the sublimation or deposition is out of our present scope. 

In pure ice or snow there are no mineral or organic particles, thus $\eta=1$. However, in ``dirty snow'', ``watermelon snow'' or in cryoconite we have $\eta <1$, and $\eta=\eta(x)$, since the top portions contain dust or microbial particles, and bottom portions contain small rocks.

The model is strongly dependent on the depth variable within the spatial variable $x$.   In particular, the temperature at the top of snow layer is close to the atmospheric temperatures which vary daily, and  at the ground surface they are close to the melting temperature $\tfr \approx 0$. The model in this paper does not describe what happens with liquid water, or that the snow may compact; these features are deferred to future work. 

\begin{remark}
\label{rem:noneq}
In snow at the microscale, the size  $\abs{ \gamma_{il}}$  of the interface region $y: \chi(x,y) \in (0,1)$ is small but nonzero. However, at the macro-scale, there might be a nontrivial interface region of $x: S_i(x)>0,S_l(x)>0$ out of equilibrium with $\theta(x)$.
While we acknowledge this difficulty, accounting for the history of the freezing and thawing is out of our scope;  see Remark~\ref{rem:EQscales}. 
\end{remark}

In the end we make a choice of $\chi$ to be of equilibrium type, and any of the choices of the generalized $\chi$ from \eqref{eq:chi} will work.   
If $\eta \approx 1$, we work with a model close to that in $\Omegaw$ and we have $
\chi= \hH(\theta-\tfr)$. For $\eta(x) \ll 1$, we choose $\chi$ as in \rass{ass:upsilon} with $\chis\ll 1$. 

The phase properties are given by
\ba
\label{eq:cksnow}
c(x)=\left[ \begin{array}{lc}
1-\eta(x) & c_r
\\
\eta(x) (1-S_g(x)) \chi & c_l
\\
\eta(x) (1-S_g(x)) (1-\chi) & c_i
\\
\eta(x) S_g(x) & c_g
\end{array}
\right], 
\; 
 k(x) = \left[ \begin{array}{cc}
1-\eta(x) & k_r
\\
\eta (1-S_g(x)) \chi & k_l
\\
\eta (1-S_g(x)) (1-\chi) & k_i
\\
\eta S_g(x) & k_g
\end{array}
\right]
.
\ea
%%%%
At each $x$, one can calculate the averages $\ave{c}^{[A]}(x)$ depending on $\eta(x),S_g(x),\chi(\theta(x))$.  

%%%%
\subsubsection{Snow (clean) under (EQ) assumption:  $S_g(x)$ known and $\mymu_g^W \approx const \neq 0$ with $\eta=1$ and $\chis=0$}
We have $S_i+S_l+S_g=1$, and we choose the (EQ) model 
\bsa
\ba 
\chi= \hH(\theta-\tfr).
\ea
%%%
With this we have for $\theta<\tboil$ 
%%%
%%
\bas
w = \left\{
\begin{array}{ll} 
c_i (1-S_g) (\theta-\tfr)+ c_g^A S_g (\theta-\tfr)
,& \theta<\tfr, \mymu_g^A=1,
\\
L_{il} \chi (1-S_g) +c_g^A S_g (\theta-\tfr) ,& \theta=\tfr,
\chi_l \in [0,1]
\\
L_{il}  (1-S_g) + c_l (1-S_g) (\theta-\tfr)+ 
c_{ave}^g S_g (\theta-\tfr) ,& \tfr < \theta <  \tboil ; \; \mymu_g^W \leq \mymu_g^{W,*} 
\end{array}
\right. 
\eas
where $c_{ave}^g= c^A_g \mymu_g^A + c_g^W \mymu_g^W $. Based on the approximations from  Section~\ref{sec:internal} we obtain a simpler expression
\ba
w(x)  = \left\{
\begin{array}{ll} 
c_i (1-S_g) (\theta-\tfr)+ c_g S_g (\theta-\tfr)
,& \theta<\tfr, 
\\
L_{il} (1-S_g) \chi ,& \theta=\tfr,
\\
L_{il}  (1-S_g) + c_l (1-S_g) (\theta-\tfr)+ 
c_g S_g (\theta-\tfr) ,& \tfr < \theta ; \; 
\end{array}
\right. 
, \; 
\ea
The state of the system is defined with the knowledge of $S_g(x)$, with $\psi=(\theta,\chi) \in \gG[\tfr,0,0]$. 
At each $x$, the left $w_*(x)$ and right $w^*(x)$ limits  of $w \vert_{\tfr^{\mp}}$ are given by 
\ba
w_*(x) = 0, \; w^*(x) = L_{il}  (1-S_g).
\ea
The inversion formula follows
\ba
w_*>w &\to&  \theta =\tfr + \frac{w}{c_i(1-S_g)+c_gS_g},  \chi =0
\\
w_* \leq w \leq w^*: w &\to&  \theta=\tfr,  \chi =   \frac{w}{L_{il}(1-S_g)}
\\
w^* < w: w &\to& 
\theta =\tfr + \frac{w - L_{il}(1-S_g)}{c_l(1-S_g)+c_g S_g},  \chi = 1.
\ea
Since $w$ is multivalued, there is no $c_{app}$ possible. 

Calculation of $k(x)$ holds since when 
$x \in \Omega_i, \chi(x) =0$ and $x \in \Omega_l, \chi(x) =1$, thus 
\ba
\label{eq:cksnowr}
 k\vert_{\Omega_i} = \left[ \begin{array}{cc}
1-\eta(x) & k_r
\\
\eta (1-S_g(x)) & k_i
\\
\eta S_g(x) & k_g
\end{array}
\right]
, 
 k\vert_{\Omega_l}  = \left[ \begin{array}{cc}
1-\eta(x) & k_r
\\
\eta (1-S_g(x)) & k_l
\\
\eta S_g(x) & k_g
\end{array}
\right]
.
\ea
%%%%
%%
Average $\ave{k\vert_{\Omega_p}}^{[G]}$  or upscaled values $\ave{k\vert_{\Omega_p}}^{[UP]}$ can be calculated for these for $p=i,l$, and for clean snow we have 
\ba
\label{eq:cksnowsimple}
 k\vert_{\Omega_i} = \left[ \begin{array}{cc}
(1-S_g(x)) & k_i
\\
S_g(x) & k_g
\end{array}
\right]
,\;
 k\vert_{\Omega_l}  = \left[ \begin{array}{cc}
(1-S_g(x)) & k_l
\\
S_g(x) & k_g
\end{array}
\right]
.
\ea
%%%%
\esa

%%%%
\subsubsection{Snow (dirty) under (EQ) assumption:  $S_g(x)$ known and $\mymu_g^W \approx const \neq 0$ with $\eta<1$ and $\chis=0$}

This model follows minor adjustments to the coefficients calculated above, consistent with the model for $\Omegas$ but with $\chi=\hH(\theta-\tfr)$. 

%%%%%%%%
\subsubsection{Cryoconite, under (EQ) assumption} 
We allow $S_g=const$ and $\mymu_g^W \approx const \neq 0$ with $\eta \ll 1$ and $\chis \geq 0$. Since cryoconite is typically located next to the surface, we expect equilibrium of liquid fraction $\chi$ with $\theta$. 
However, due to nontrivial presence of rock or organic particles, we expect that some portions of water at the micro-scale  freeze at temperatures other than $\tfr=0$. 

For these we postulate a general model from Section~\ref{sec:soils}, since the properties may be similar to $\Omegas$.

For large $\eta$, the model is similar to that in ``dirty snow'' in $\Omegasn$.  

%%%%%%%%%%%%%
\section{Examples of effective data}
\label{sec:examples}

The most general model \eqref{eq:defPs} from Section~\ref{sec:soils} allows $\eta \leq 1 $, $\chis \in [0,1]$, and a fixed $S_g\geq 0$, and allows either multivalued or single valued piecewise smooth $\chi(\theta)$. In this section we provide examples and illustrations useful for the future numerical approximations.

First we focus on the data $\chieff,\keff$ which now account for the presence of air pockets and a wide collection of pore sizes including the macro-pores. To calculate $\chieff$, we only require average $\tfr$ in each of the nano-, micro- meso-, and macro-pores, along with their volume fractions.  In section~\ref{sec:chiexample} we explain how to obtain $\chieff$ and fit to $\wt{\chieff}$ starting from an assumed distribution of $\tfrm$
based on the conceptual cartoon in Figure~\ref{fig:porousm}.
In Section~\ref{sec:NUSFF} we use computations instead of a cartoon, and guide how to find $\omegai,\omegag$ as well as how to assign $\tfr$. In Section~\ref{sec:ccexample} we provide  closed form calculations of $\wt{\cCeff}$ calculated for $\wt{\chieff}$ from Section~\ref{sec:chiexample}.  
Finally, in Section~\ref{sec:kexample}  we calculate the effective conductivity $\keff\approx \ave{k}^{[G]}$ for $\Omegasn$ and compare to the empirical correlations from literature we discussed in Section~\ref{sec:snowlit}. 
%%%
\subsection{Finding $\chieff$ and $\wt{\chieff}$ when macro-pores are present} 
\label{sec:chiexample}

We consider now two examples motivated by the distribution of nano-, micro-, meso-, and macro-pores from a cartoon shown in Figure~\ref{fig:porousm}.
\begin{example}
\label{ex:chiexample} 
 We estimate the pore sizes $(r^{(m)})_m$ from the cartoon for four subdomains and estimate $\tfrm$ from 
 Gibbs-Thomson law applied to calculate $\tfrm \sim 1/r^{(m)}$ as suggested in \cite{PVB}. We obtain 
 $\tfrm = (-15,-5,-1,0)$.
We also estimate the volume fractions of these domains, assuming in case (A) $\etam=
(0.1,0.35,0.50,0.05)$ and case (B) that  $\etam=
(0.1,0.2,0.40,0.30)$. For these $\chieff$ are piecewise constant functions. We find next a fit of each to some  $\Upsilon(\theta)=\chis e^{b\theta}$. We find that $\chis=1,b=0.2$ works for case (A), but $\chis <1,b=0.2$ is needed for case (B). The plots of $\chieff$ and $\wt{\chieff}$ are given in Figure~\ref{fig:chi}.
\end{example} 

%%%
\subsection{Creating synthetic  distributions of $\omega(x)=\omegai \ccup \omegai \ccup \omegag$}
\label{sec:NUSFF}

In order to calculate $\chieff, \cCeff$ in \eqref{eq:ccdefPs} one needs to know $S_g$ as well as $\tfrm$. In order to find $\keff$, one needs a large enough representative sample of realistic distributions of actual phases $p=r,l,i,g$ at the pore-scale. Since imaging data is limited, we create synthetic data sets, and study which surrogate models for $\keff$ are adequate.

As described in Example~\ref{ex:chiexample}, to get $\chieff$, we require a broad collection of $\tfr^{(m)}$ and $\eta^{(m)}$. While the use of Gibbs-Thomson is intuitively clear and easy to apply for a proof-of-concept case, it appears insufficient to explain the range of $\tfr$ based only  on the pore sizes, e.g., obtained from imaging \cite{Rooney22}; see discussion in \cite{PVB}. At the same time, Direct Numerical Simulations (DNS) simulations  of \eqref{eq:heat} at the pore-scale carried out in \cite{PHV}  to get $\omegai$ are possible, but they rely on the boundary and initial data which are actually unknown. This presents a conundrum.

\begin{figure}[htp]
\centering
\includegraphics[height=3cm]{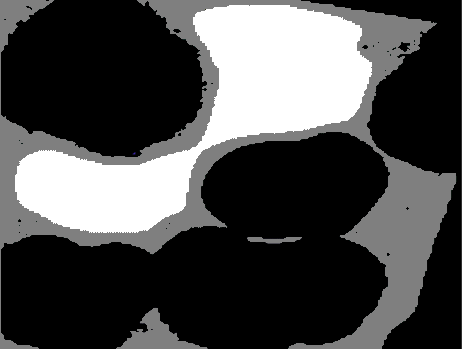}
\includegraphics[height=3cm]{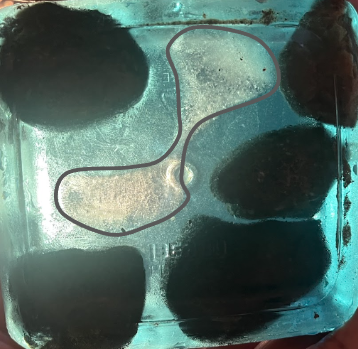}
\caption{Computational and experimental efforts from Example~\ref{ex:NUS} illustrating the formation of ice $\omega_i$ in a pore-scale domain $\omega(x)$. Image on the left comes from simulation with resolution $500 \times 500$ of the landscape equation (left), with $\omegar$ in black, ice $\omegai$ in white, and liquid water in $\omegal$ in gray. Photo on the right is of ice formation.  }
\label{fig:NUS}
\end{figure}

\begin{example}(Distribution of $\omegai \subset \omega(x)$)
\label{ex:NUS}
 One approach to study an average $S_i$ in a pore is to postulate (*) that the ice domain $\omegai$ forms preferentially away from the pore boundaries at $y: \text{dist}(y,\omegar)$ is large. To quantify this distance from $\omegar$, we use an auxiliary singularly perturbed elliptic PDE $-\mu \Delta u +u=1$ with homogeneous Dirichlet boundary conditions on $\partial \omegar$, also called the ``landscape equation'', Since $u \approx 1$ away from $\partial \omegar$, we can define $\omega_i=\{y: \abs{u(y)-1}\leq tol\}$ to some desired tolerance. The parameters $tol$ and $\mu$ can be adjusted depending on the desired $S_i$.  To find $u(y)$, we use the Finite Element Method; specifically, we adapted the code in \cite{ACF} and set up meshing of the domain $\omega\setminus \omegar$.  In addition, we set up a physical experiment in a domain of $diam(\omega) \approx 10~\mpunit{cm}$ to check if indeed (*) holds.  The comparison of these two is very satisfactory, as shown in Figure~\ref{fig:NUS}. 
\end{example} 
We aim to extend this approach in the future once more physical data on the interactions of ice phase with air is available.  

Meanwhile, we study the impact of the distribution of $\omegal, \omegai, \omegag$ on $\ceff,\keff$. To this aim, we can use xray $\mu$-CT data such as in \cite{calmels}. 
However, we need a good number of such domains for bona-fide testing. We generate these using the MCMC (Markov Chain Monte Carlo) Metropolis algorithm from statistical mechanics. Given $\omegar \subset \omega$ from xray $\mu$-CT, we predict the position of subdomains occupied by different phases based on postulated information about their aggregation. For this, we adapt the MCMC variant we developed for hydrate and biofilm subdomains in \cite{PUS}, and extend it to the present context of $\omegal,\omegai,\omegag$. This requires setting up a Hamiltonian for the (aggregating) interactions $pp$ for all $p$ and $rl$, and (repelling) interactions for $ri,rg$, and neutral for other pairs. At this time, these are found by trial and error, but we plan to find these by machine learning in the future. We give details in Example~\ref{ex:FF}. 

%%%
\begin{table}[ht]
{\scriptsize
\begin{tabular}{|c|ll|llll|l|lll|}
\hline
&
$\eta$
& 
$\eta_g$
&
$S_l$ 
&
$S_i$
&
$S_a$ 
&
$\chi$
&
$\ave{c}^{[A]}$    
&
$\ave{k}^{[A]}$ 
&
$\ave{k}^{[G]}$ 
&
$\ave{k}^{[H]}$
\\
\hline
%%%%%%%%%%%%%%%%%%%%%%%%%%%%%%%%%%%%%%%%%%%%%%%%%%
(A)
&
$0.443$& 
$0.0221$
&
$0.475$ 
&
$0.474$
&
$0.049$ 
&
$0.500$
&
$1.92$    
&
$0.77$ 
&
$0.50$ 
&
$0.31$
\\
soil
&&&&&&&&\multicolumn{3}{|c|}{
$\ave{k}^{[UP]}=[0.476,0.00067, 0.00067, 0.480]$}
\\
%\hline
100$\times$100
&
$0.443$& 
$0.0$
&
$0.525$ 
&
$0.474$
&
$0$ 
&
$0.525$
&
$2.014$    
&
$0.785$ 
&
$0.5368$ 
&
$0.425$
\\
&&&&&&&&\multicolumn{3}{|c|}{
$\ave{k}^{[UP]}=[0.511, 0.001, 0.001, 0.516]$}
\\
\hline
(B)
&
0.555
&
0.0277
&
0.475
&
0.475
&
0.05
&
0.5
&
2.121
&
0.893
&
0.570
&
0.320
\\
soil
&&&&&&&&\multicolumn{3}{|c|}{
$\ave{k}^{[UP]}$ unavailable}
\\
1000$\times$500
&
0.555
&
0
&
0.525
&
0.475
&
0
&
0.525
&
2.237
&
0.908
&
0.621
&
0.476
\\
&&&&&&&&\multicolumn{3}{|c|}{
$\ave{k}^{[UP]}$ unavailable}
\\
\hline
(C)&
1 & 0.421
& 0.022 & 0.557 & 0.421 & 0.038
& 1.179 & 1.30 & 0.338 & 0.0607
\\
snow
&&&&&&&&\multicolumn{3}{|c|}{
$\ave{k}^{[UP]}=[0.2107, 0.005, 0.005, 0.2924]$}
\\
\hline
(D)&
0.978 & 0.421
&0 & 0.569 & 0.430 
& 0&
1.11 & 1.298 & 0.333 &0.061
\\
cryoconite
&&&&&&&&\multicolumn{3}{|c|}{
$\ave{k}^{[UP]}=
[0.2096, 0.0051, 0.0051, 0.2914]
$}
\\
\hline
\end{tabular}
}
\caption{Results of Example~\ref{ex:FF}. For (B), the upscaled $\ave{k}^{[UP]}$ is not available due to size of the data set.}
\label{tab:FF}
\end{table}

\begin{figure}[htp]
\centering
\includegraphics[height=4cm,angle=-180]{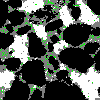}
\includegraphics[height=4cm,angle=-180]{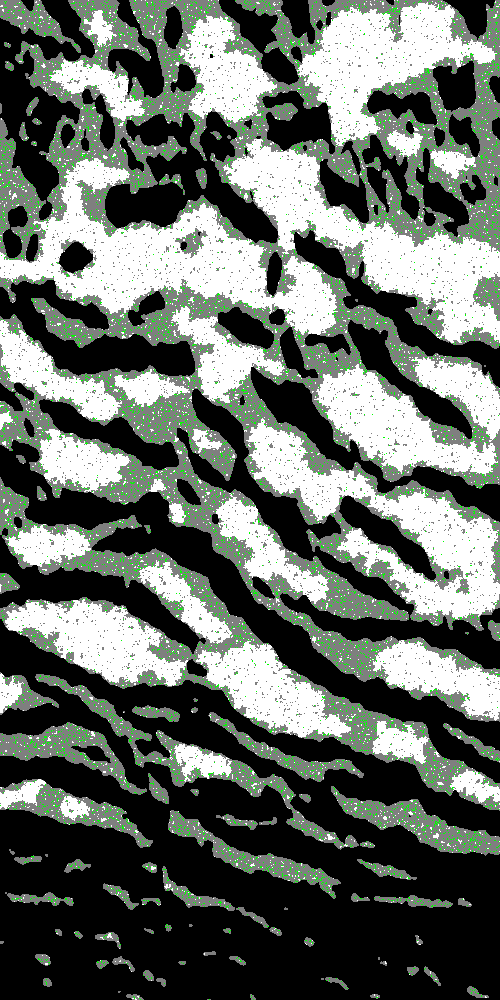}
\centerline{\hfill (A) \hfill (B) \hfill}
\caption{Illustration of heterogeneous domain $\omega(x)$ from Example~\ref{ex:FF} generated by MCMC algorithm for cases (A) and (B). Distribution of $\omegar$ (black), $\omegal$ (gray), $\omegai$ (white), $\omegaa$  (green).}
\label{fig:FF}
\end{figure}

\begin{figure}[htp]
\centering
\includegraphics[height=4cm]{
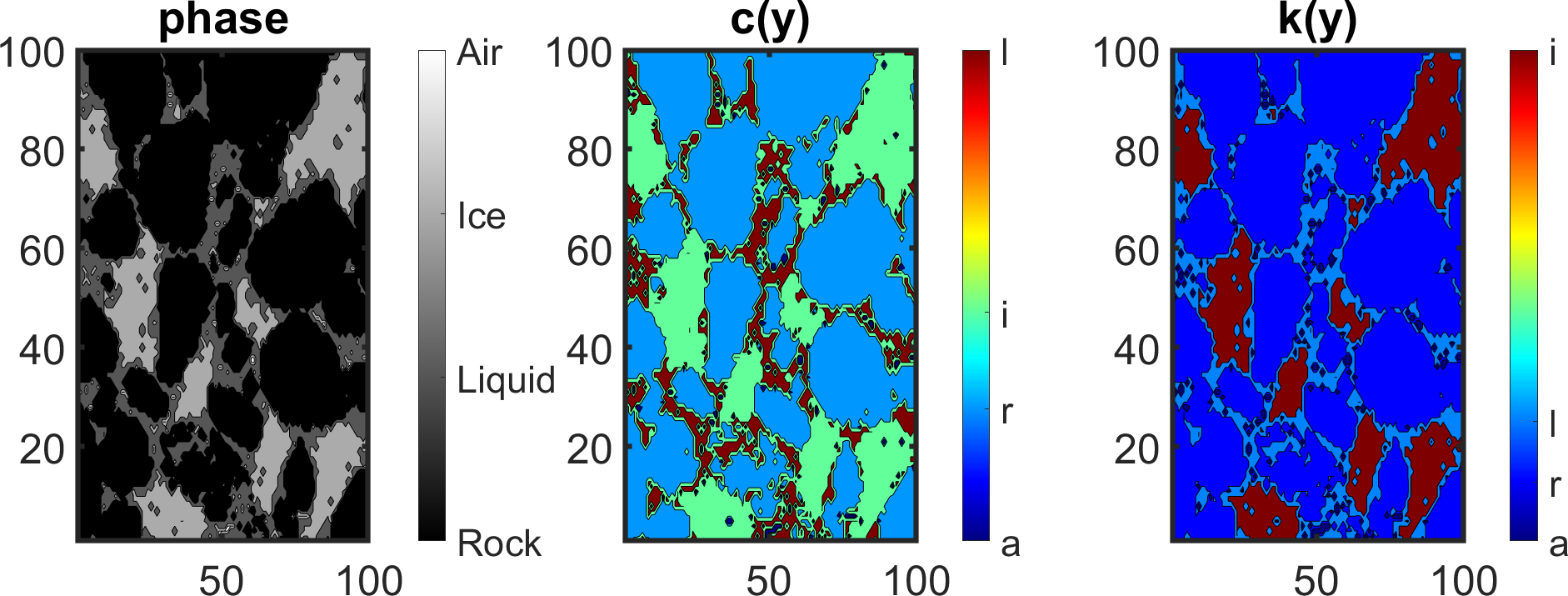}
\\
\includegraphics[height=4cm]{
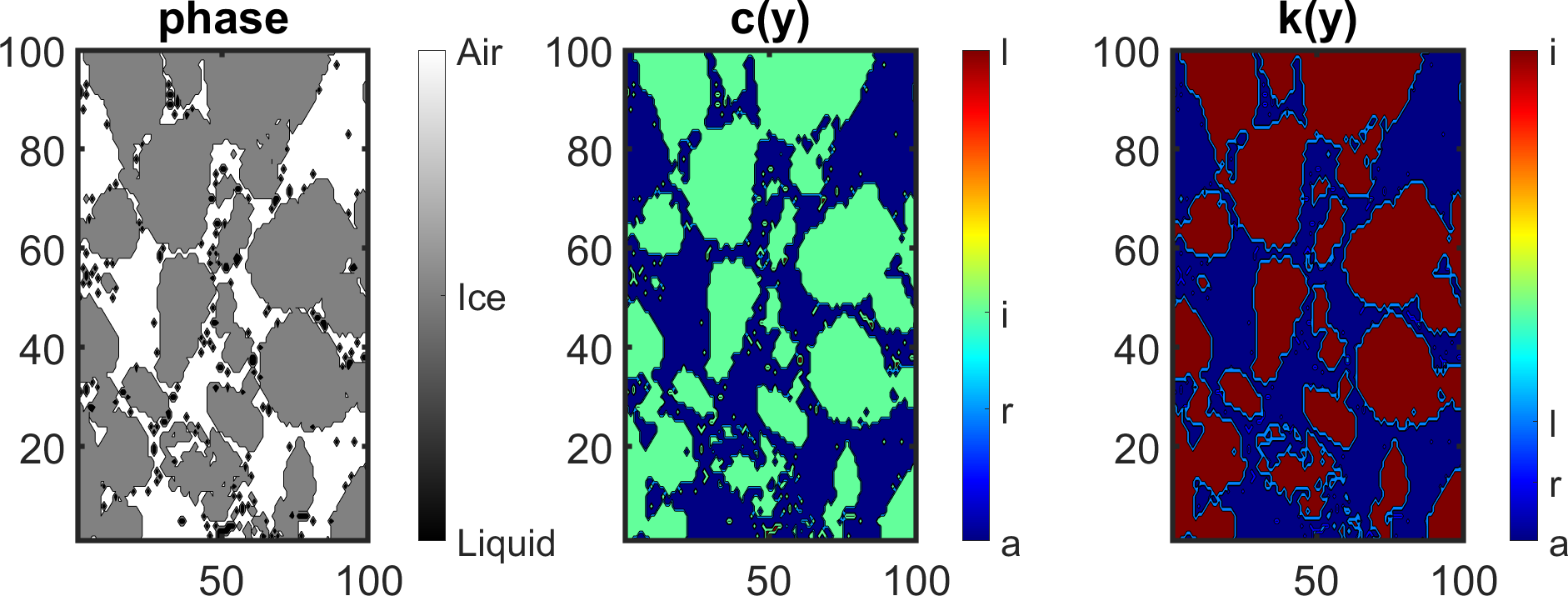}
\\
\includegraphics[height=4cm]{
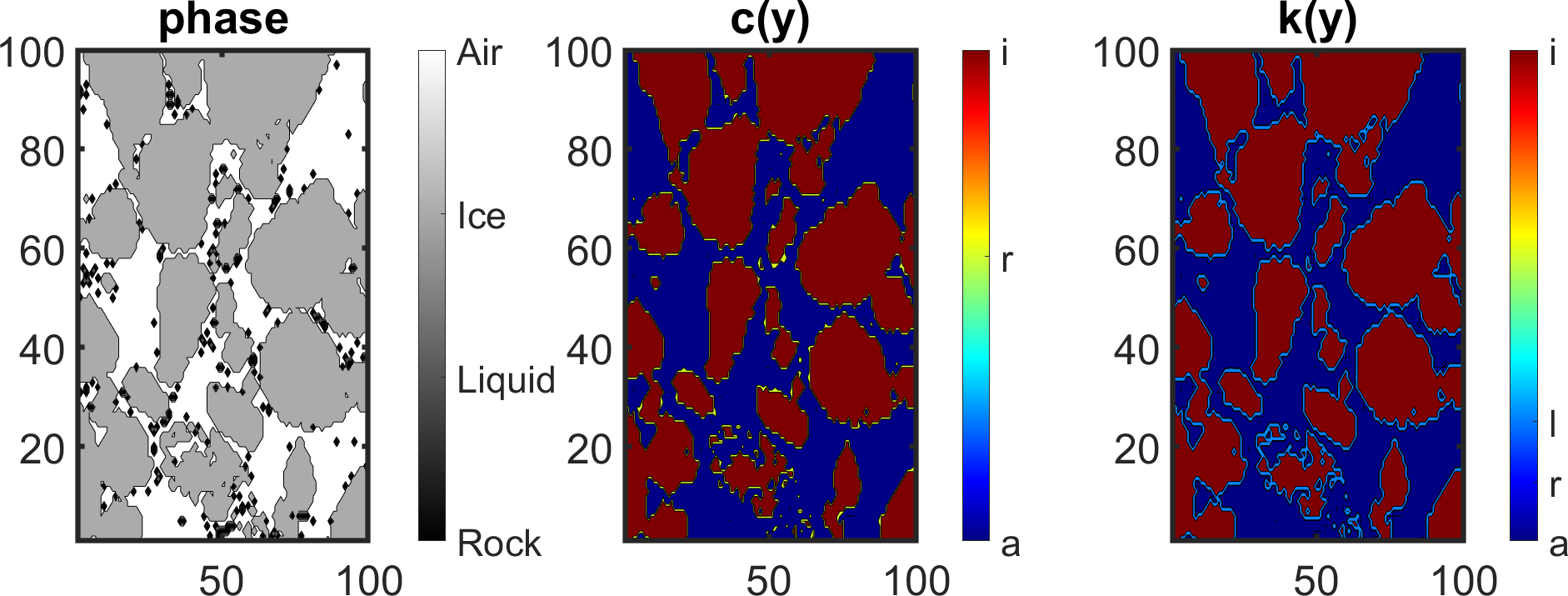}
\caption{Illustration of heterogeneous properties for $ y \in \omega(x)$ from Example~\ref{ex:FF}. From left to right: phases $\omegap,p=r,l,i,g$, pointwise $c=c(y)$ and $k=k(y)$.   Top: soil with $S_g>0$. Middle: snow with large $S_g,S_i$, and $\eta=1$, $S_l \approx 0$. Bottom: cryoconite with large $S_g,S_i$, $\eta \approx 1$, and $S_l=0$.  }
\label{fig:FFMP}
\end{figure}

%%%%%%%%
\begin{example}[Synthetic distribution of phases $i,l,g$ in $\omega$ with Metropolis and Potts models \cite{ShonkwilerMendivil,Madras2002} and upscaling to $\ceff,\keff$] 
\label{ex:FF} 
We start with images obtained by xray $\mu$-CT, and generate images with $S_g>0$, and $S_g=0$ for testing. We consider geometry (A) $100\times 100$ originally from \cite{PTISW}, and (B) geometry $1000 \times 500$ adapted from \cite{calmels} and \cite{PHV}. We apply the MCMC algorithm; for (A) it takes about $3 \times 10^6$ iterations to converge, order of 1~min wall clock time, and for (B) it takes $3 \times 10^8$, about 2~h wall clock time. For (A-B), these take $2$-$3min$. For comparison we also include the case where there is no air where we set $\omegal \ccup \omegag \to \omegal$; this case for (A) and (B) is indicated by $\eta_g=0$.  The results are in Figure~\ref{fig:FF}.  Next we apply averaging and upscaling to obtain $\ceff,\keff$ reported in Table~\ref{tab:FF}.  Unfortunately, upscaling the data set $1000\times 500$ is not possible due to the size.
\end{example}

We see from Table~\ref{tab:FF} small effect of neglecting air presence on either $\ceff,\keff$ for sets (A-B), and that the upscaled $\keff$ are very similar to $\ave{k}^{[G]}$. 

We create next a synthetic data snow set (C) based on (A) where we swap the voxel assignments $r\to i,l\to a, i \to a, a \to l$, and a cryoconite (dirty snow) data set based on (A)  where we swap the voxel assignments $r\to i,l\to a, i \to a, a \to r$. 
We illustrate these cases by plotting $\omega(x)$ as well as pointwise $c,k$ in Figure~\ref{fig:FFMP}. We continue with upscaling to $\ceff,\keff$. We see a small but insignificant difference between $\ceff,\keff$ for (C) and (D).

%%%%%%%%%%%%%%%%%%%
\subsection{Finding  $\wt{\cCeff}$ when air and macro-pores are present} 
\label{sec:ccexample}
\bsa
We first let $S_g>0$ be given, and we calculate $\cuf, \cfr$ from \eqref{eq:cufr}.  For example, when $Sg =0.1, \eta=0.5, \cfr=1.43; \cuf=2.42$. But when $Sg =0.5, \eta=0.95, \cfr=1.12, \cuf=1.18$. 

Next we  choose some $\wt{\chieff}$. We will assume with $\chis \in [0,1]$ and
\ba
\label{eq:chiM}
\Upsilon(\vartheta)=\chis e^{b \vartheta}, \vartheta\leq 0
\ea
%%%
from Example~\ref{ex:chiexample} which satisfies \rass{ass:upsilon}. If $\chi^*=1$ we recover \eqref{eq:chiP}, and if $\chis=0$, we recover \eqref{eq:chi}.  Now we follow Section~\ref{sec:soils} and generalize Example~\ref{eq:chiPs}. 

\begin{example}
\label{ex:wexample}
[Finding $\wt{\cCeff}$ for soil model $\eta<1, \chis \leq 1$ and $S_g>0$]
We have 
%%%
\ba
\label{eq:ccdefPs}
\wt{\cCeff}(\theta) = 
\left\{ \begin{array}{ll}
\cfr(\theta-\tfr) + (\cuf-\cfr)\chis  \frac{e^{b(\theta-\tfr)}-1}{b}, &\theta \leq \tfr,
\\
\cuf(\theta-\tfr), &\theta>\tfr.
\end{array}
\right.
\ea
We see that $\wt{\cCeff}(\theta)$ is continuous on $\R$. Its derivative is nonnegative on $\R_-\cup \R_+$, and takes a jump at $\theta=\tfr$, with the left limit $\cfr (1-\chis) + \cuf \chis$ and the right limit $\cuf$.  We complete the full definition of $w$ adding $L\eta \chi$ with $\chi$ given from \eqref{eq:chidefP} as follows
\ba
\label{eq:enthalpydefPs}
w = \left\{ \begin{array}{ll}
\cfr (\theta-\tfr) + (\cuf-\cfr) \chis \frac{e^{b(\theta-\tfr)}-1}{b} + L \eta \chis e^{b(\theta-\tfr)}, &\theta<\tfr,
\\
{[L\eta\chis,L\eta]}, &\theta=\tfr,
\\
\cuf (\theta-\tfr) +L\eta, &\theta>\tfr,
\end{array}
\right.
\ea
Unless $\chis=1$, we cannot calculate $c_{app}$ and cannot calculate $\partial_t w$. However, since $\cuf>\cfr$, it is easy to see that each summand of the expression for $\theta<\tfr$ is monotone increasing, 

If $\chis=1$, the function is monotone increasing, and a bijection, with the limiting behavior dominated by the linear term with slope $\cuf$ ad $\cfr$, respectively. 
\end{example} 

The relationship $\theta \to w$ is set valued but on the monotone graph $\gG[\tfr,\chis,\Upsilon] \ni (\theta,\chi) \to w$ is a continuous piecewise smooth bijection. The limits  at $\tfr^{\mp}$ are $w_*=L\eta \chis,w^*=L\eta$, respectively.

Regarding inversion of \eqref{eq:enthalpydefPs}, it is trivial for $w>w^*$. For $w \in (w_*,w^*)$ we get $\theta=\tfr,\chi= \frac{w}{L\eta}$. For $w<w_*$, upon substituting $\vartheta=\theta-\tfr$, we seek 
%%%
\ba
\label{eq:localN}
\vartheta: w=g(\vartheta).
\ea
\esa
where
\ba
\label{eq:gdef}
g(\vartheta)=\alpha \vartheta + \beta e^{b\vartheta} + \gamma
\ea
where $\alpha=\cfr$, $\beta=\chis \frac{\cuf-\cfr}{b} + L \chis \eta$, and $\gamma=-\chis \frac{\cuf-\cfr}{b}\leq 0$.  Now $g(\cdot)$ does not have a closed form algebraic inverse $g^{-1}$; one might be tempted to use some transcendental formulations, but a robust approach is to develop local nonlinear solver.
Since $\lim_{\vartheta \to -\infty}g(\vartheta)=-\infty $ and  $\lim_{\vartheta \to 0^-}g(\vartheta)= w_*=L\eta \chis$, its zero is bracketed on $\R_-$. We outline the use of some nonlinear solvers in Section~\ref{sec:nsolver}. 

\begin{figure}[htp]
\centering
\includegraphics[height=4.5cm]{
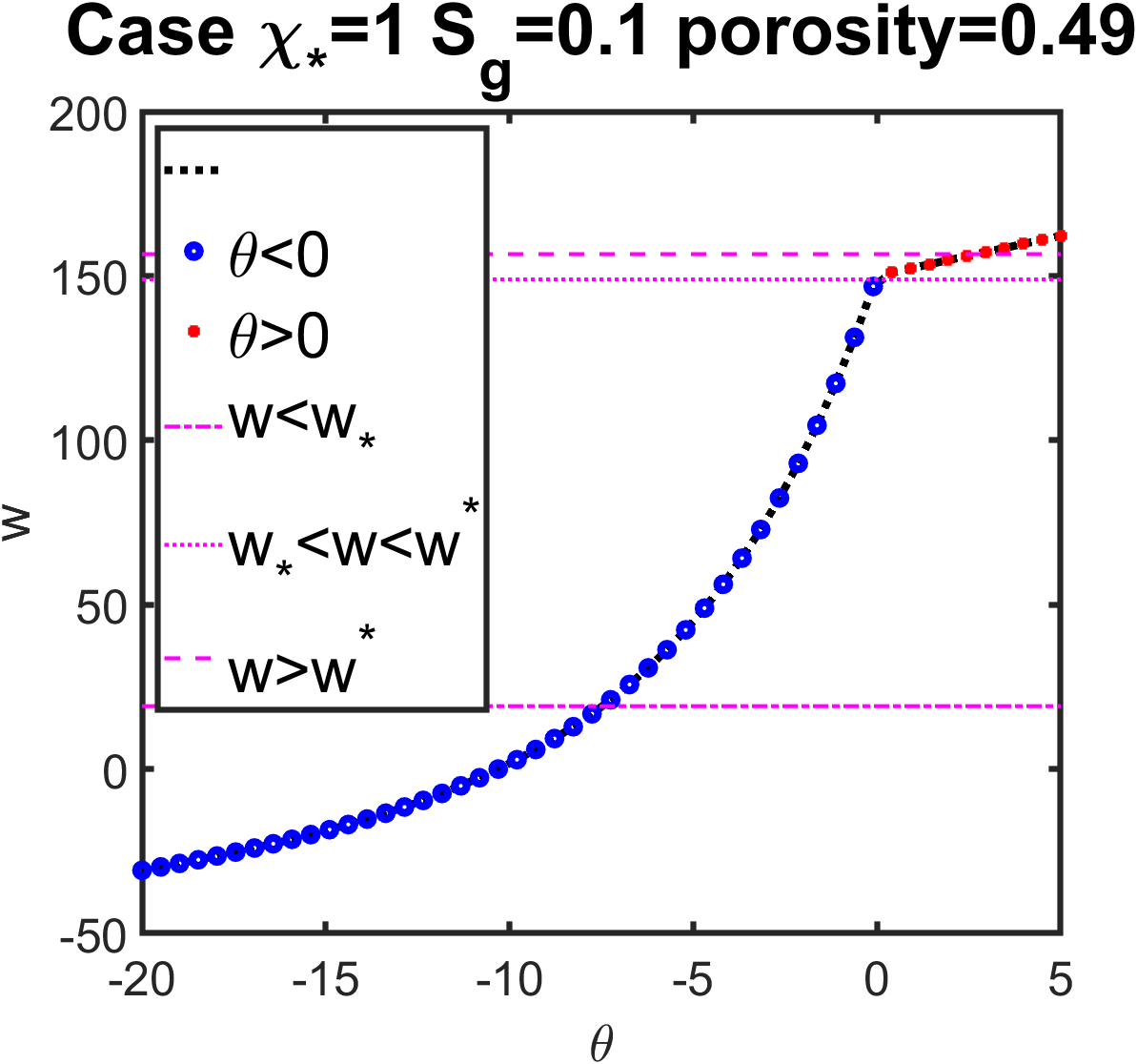}
\includegraphics[height=4.5cm]{
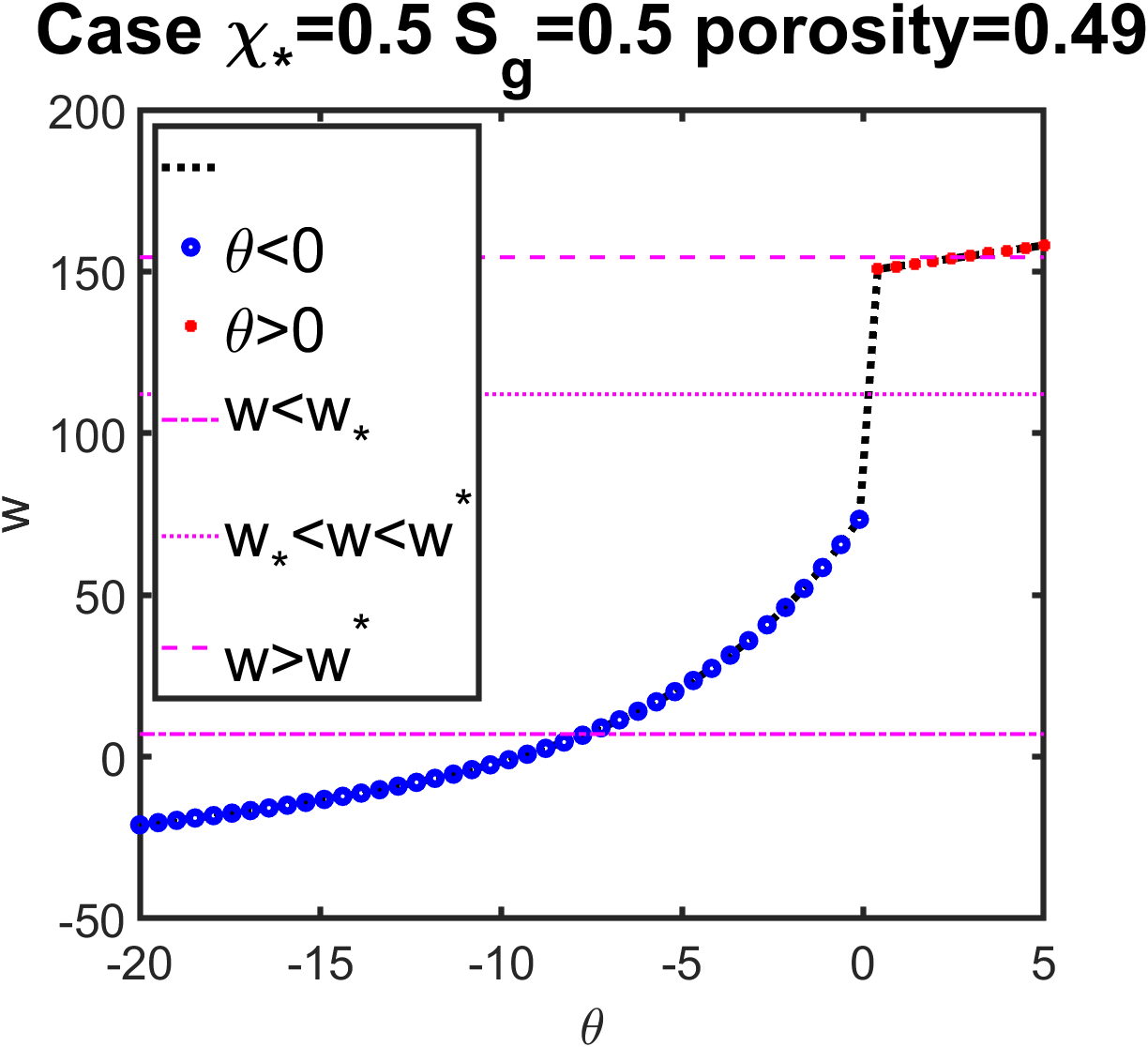}
\includegraphics[height=4.5cm]{
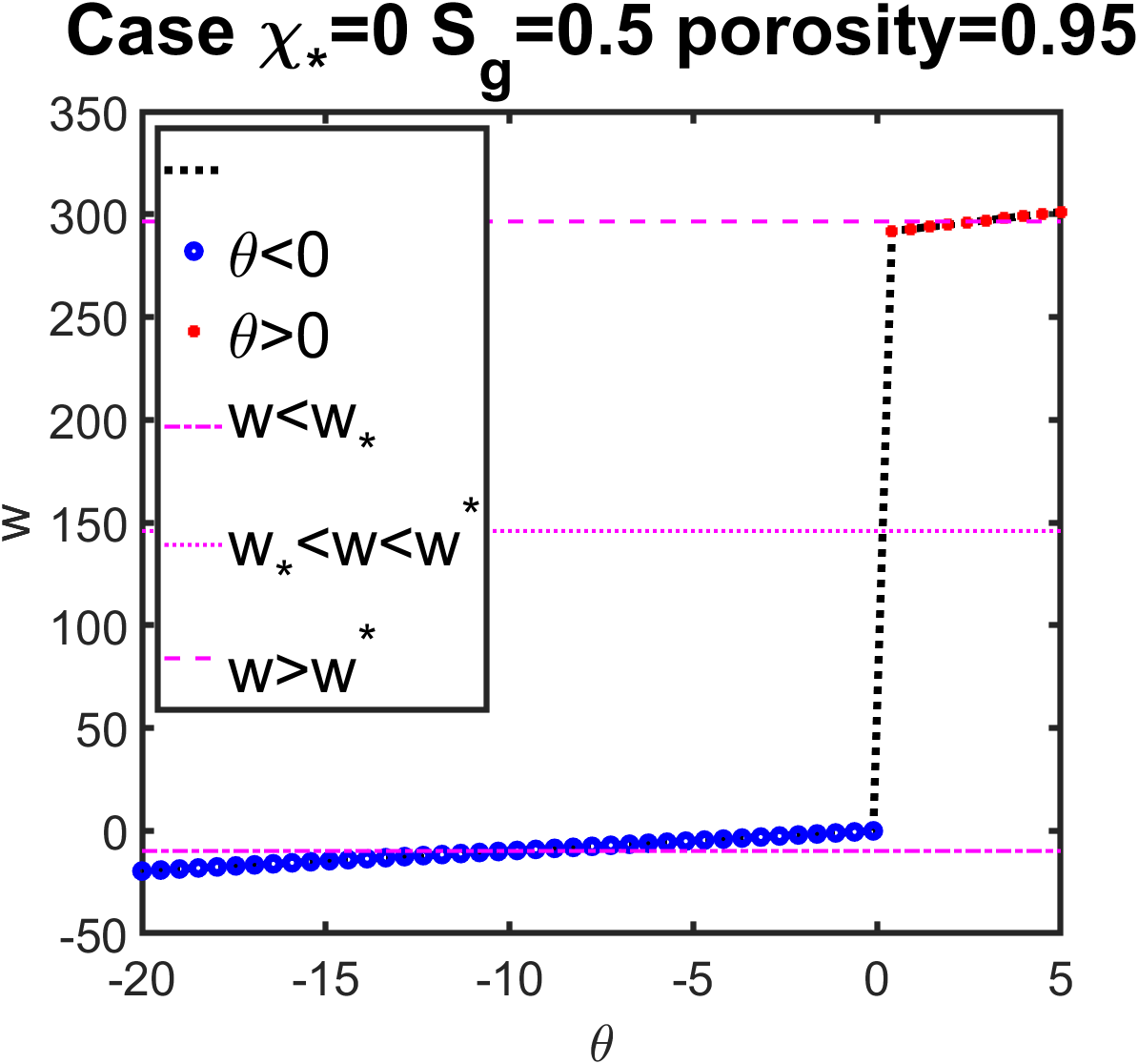}
\caption{Illustration of the general formulas $(\theta,\chi) \to w$  and constructing inverses  from Example~\ref{ex:wexample}, with data $\chis,S_g,\eta$ as indicated on each plot. From left to right, these cases are typical for (left) soil without macro-pores but with air, (middle) soil with macro-pores or cryoconite, and  (right) ``dirty'' snow.}
\label{fig:wexample}
\end{figure}

We plot examples of the application of 
\eqref{eq:enthalpydefPs} in Figure~\ref{fig:wexample}. For each, we plot $w$ as well as three cases requiring inversion: $w<w_*,w_*<w<w^*,w^*<w$. For the latter two which are multi-valued, finding $\theta$ does not describe the state of the problem, and we mist additionally calculate $\chi$. 

%%%%%%%%%%%%%
\subsection{Thermal conductivity of snow and cryoconite}
\label{sec:kexample}
In Section~\ref{sec:snowlit} we indicated that in the literature, discussions of the snow conductivity seem to not provide consensus; see, e.g., \cite{LingZhang,Sturm2002,Jafarov2014}. On the other hand, we can calculate an upscaled value of $\keff$ with the formulas from Section~\ref{sec:keff}.   Since we do not have geometry of snow microstructure other than the hypothetical geometry in Figure~\ref{fig:FFMP}, and the geometry is required for 
$\ave{k}^{[UP]}$, we can estimate $\keff \approx \ave{k}^{[G]}$.  

To do so, e.g., when $k$ is represented as in the formula \eqref{eq:cksnow} we need $k_i,k_g,S_i,S_g$ as well as $k_l$ whenever $S_l \neq 0$.  However, to compare to the correlations of $k$ depending on the density $\rho^{snow}$ in \cite{LingZhang,Jafarov2014,Sturm2002}, we must calculate this density depending on $\rho_g,\rho_i,\rho_l$ and the known phase fractions. 
This approach turns out to be promising. 

\begin{figure}[htp]
\centerline{
\includegraphics[height=2in]{
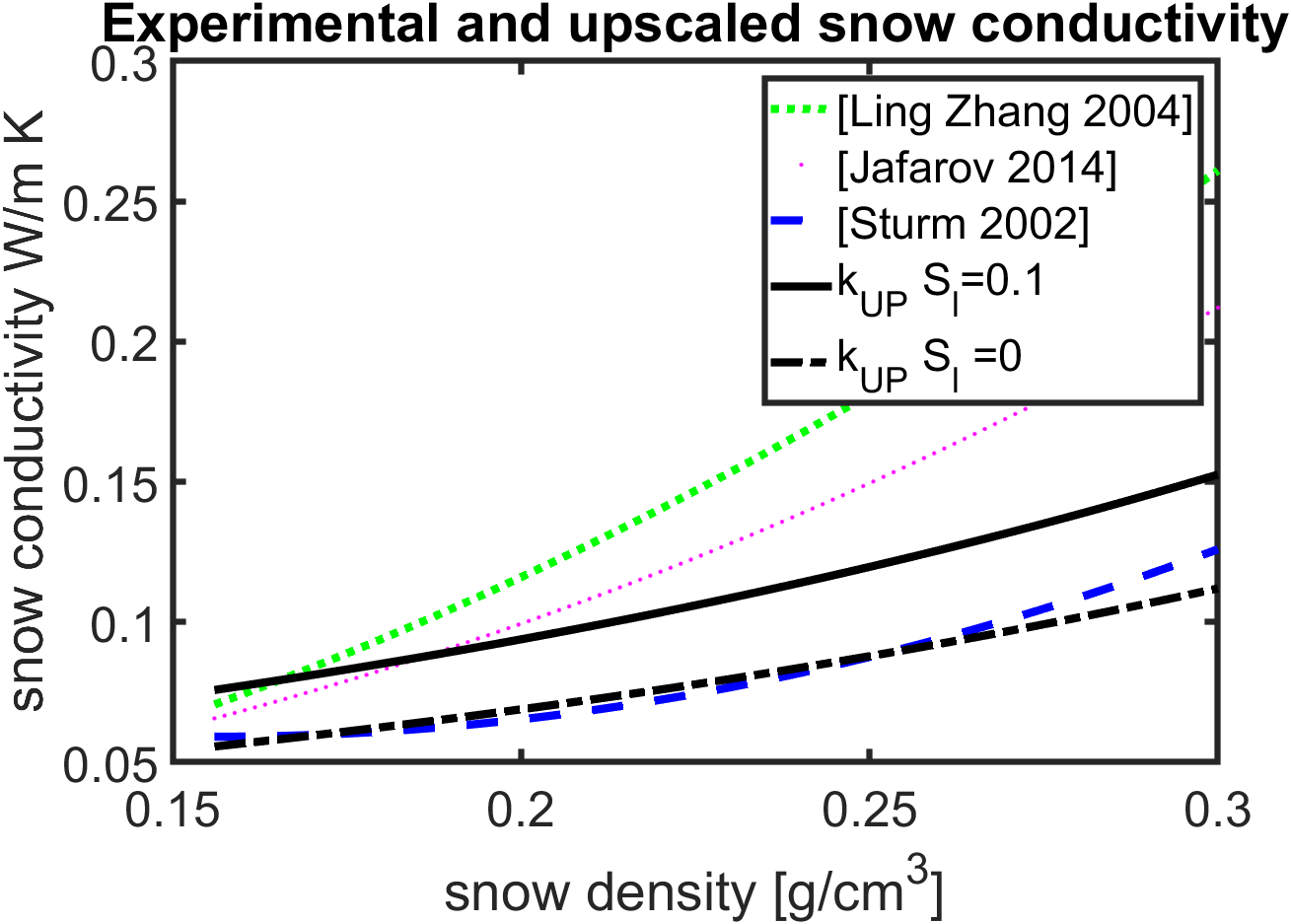}
}
\caption{Results of Example~\ref{ex:thermal_lambda}: comparison of snow conductivity obtained from empirical formulas in \cite{LingZhang,Jafarov2014,Sturm2002} with the upscaled $\keff$ we derive.}
\label{fig:ksnow}
\end{figure}

%%%%%%%%%%%%%
\begin{example}[Comparison to empirical formula of effective coefficients] 
\label{ex:thermal_lambda}
We aim to compare the formulas from \cite{LingZhang, Jafarov2014, Sturm2002}.  Assume the snow density $\rho^{snow}$ is known. Then assume a given constant $S_l$, and  $S_g+S_i =1-S_l$, thus we can find $S_i = \frac{\rho^{snow}-\rho_a -S_l\rho_l}{\rho_i-\rho_a}$, from which $S_g$ follows. For each we calculate now $\ave{k}^{[G]}$ for $k$ given for snow as composite in \eqref{eq:cksnow}. 

With these calculations we can now compare $\ave{k}^{[G]}$ depending on $\rho^{snow}$ to the literature. Specifically, we have \mpcitee{LingZhang}{eq. 2.9} formula which gives 
$k_{LZ} = 2.9 (\rho^{snow})^2$. We also have 
\bas
k_{JAF} = 2.5(\rho^{snow})^2 - 0.123 \rho^{snow} + 0.024
\eas
from \mpcitee{Jafarov2014}{eq. 6}, and $k_{STURM} =    0.138  - 1.01 \rho^{snow} +  3.233 (\rho^{snow})^2$ from  \cite{Sturm2002}. We plot these three relationships against that we found $\ave{k}^{[G]}$ through inverting to get $S_i$. See Figure~\ref{fig:ksnow}. 
\end{example}

We find discrepancy between these three models themselves but also that our upscaled formula matches these quite well at least for small $\rho^{snow}$.

%%%%%%%%%%%
\section{Summary and conclusions}

In this paper we extended the framework  previously applied to air-free soils for modeling thermal conduction with phase transition from micro- to macro-scale.   The new models derived here apply to soils with air pockets and very large pores, to the snow in some conditions, and to cryoconite. The models depend on parameters $\chis \in [0,1]$ and $S_g$ which are assumed known. Along with other physical constants, we obtain a very useful general relationship $\theta \to w$. 

This new unified modeling framework provides an explanation and clarification on the data needed for modeling. In addition, with the application of the concept of multiple materials, it allows to create monolithic numerical models useful for modeling of coupled soil and snow models, or of cryoconite.  

Since the relationship $\theta \to w$ is multi-valued when $\chis <1$, one has to incorporate properly the liquid phase fraction $\chi$. With this, we show that $(\theta,\chi) \leftrightarrow w$ is a bijection.  A critical element of the general model is an ability to invert the relationship $w \leftrightarrow \psi = (\theta,\chi)$, where $(\theta,\chi)$ lie on a single parameter parametrized curve in $\R^2$. 

In the paper we provide mathematical framework, and illustrate with examples which show useful applications of this framework. 

%%%%%%%%%%%%%
\section{Acknowledgments}
The authors wish to thank colleagues who were involved in the discussions as well as earlier steps of this research: Naren Vohra, Lisa Bigler, Zachary Hilliard, Corbin Savich, and Hannah Dempsey. This research was partially supported by the grants NSF DMS-1912938 ``Modeling with Constraints and Phase Transitions in Porous Media'',  NSF DMS-2309682 ``Computational mathematics of Arctic processes'',  and by Joel Davis faculty scholar position, PI: Malgorzata Peszynska. Dempsey, Felsch and Unger-Schulz were partially supported by the Oregon State University URSA program. 

We also wish to thank
the anonymous reviewers for their remarks which helped to improve the exposition in the paper. 

\section{Appendix}
%%%%%%%%%
\subsection{Nonlinear solver to find the root of \eqref{eq:localN}}
\label{sec:nsolver}
Given $w: w<\ws$, the residual of $g(\vartheta)-w=0$ is bracketed in $(-\infty,0]$. We recall that a (continuous) function is bracketed on some interval $[a,b]$ if it is has a nonpositive value at the left end, and nonnegative value at the right end. Extending this notion to the case when $a \to -\infty$ requires that the value is replaced by the limit which should be strictly negative. 

Therefore, a minor modification of  the bisection method will easily find the root of \eqref{eq:localN}. First (STEP 0) we find some $\vl:$ $g(\vl)-w<0$: this is easy to do starting with $\vl=0$ and iterating $\vl :\to \vl-1$; this process must complete because $\lim_{\vartheta \to -\infty} g(\vartheta)-w=-\infty$ and  $g(\vartheta)$ is continuous. We also set $\vr=0$. Once this is complete, we use (STEP 1) and bisection in $[\vl,\vr]$ to identify $\vartheta_0: \abs{g(\vartheta_0)-w}<tol$, where $tol$ is the desired tolerance. 

Another algorithm pre-calculates a table $(\vartheta_k,g(\vartheta_k))_k$ for some large number $K$, with $\vartheta_k \in (-\infty,0]$ and replaces (STEP 0) by finding the particular $k$ for which $g(\vartheta_k) \leq w \leq g(\vartheta_{k+1})$. This process is known as an inverse function lookup, available in most software environments, and is guaranteed to succeed because $g(\cdot)$ is monotone increasing. Then we can replace (STEP 1) by Newton's algorithm:  since $g(\cdot)-w$ is smooth on $\vartheta_k,\vartheta_{k+1}$, we can apply Newton's method with line-search, in which we keep all guesses bracketed in $\vartheta_k,\vartheta_{k+1}$. The standard assumptions for convergence of Newton's algorithm \cite{kelley} apply since $g'(\vartheta) =\alpha + \beta b e^{b \vartheta} $ is always nonnegative and bounded in $[\alpha,\alpha + b \beta]$ on $\R_-$. Once the algorithm completes with some $\vartheta_0$, we obtain $\theta=\tfr + \vartheta_0$.

\section{Conflict of interest statement}
On behalf of all authors, the corresponding author states that there is no conflict of interest.

\bibliographystyle{plain}
\bibliography{PMPFUS}
\end{document}